\documentclass[fleqn,usenatbib]{mnras}

\usepackage{prettyref}
\usepackage{float}
\usepackage{rotfloat}
\usepackage{amsmath}
\usepackage{amssymb}
\usepackage{graphicx}
\usepackage{color}
\usepackage{url}
\usepackage{comment}
\usepackage{natbib}

\usepackage[T1]{fontenc}
\usepackage{ae,aecompl}

\usepackage{tablefootnote}

\makeatletter
\providecommand{\tabularnewline}{\\}

\usepackage{indentfirst}
\usepackage{leftidx}

\graphicspath{{figures/}}
\makeatother

\title[Polarization from Radiative Alignment]{Does HL Tau Disk Polarization in ALMA Band 3 Come from Radiatively Aligned Grains?}
\author[H. Yang et al.]{
Haifeng Yang$^{1,2}$\thanks{E-mail: yanghaifeng@tsinghua.edu.cn}\thanks{C. N. Yang Junior Fellow},
Zhi-Yun Li$^{1}$,
Ian W. Stephens$^{3}$,
Akimasa Kataoka$^{4}$,
\and
and Leslie Looney$^{5}$
\\
$^{1}$Astronomy Department, University of Virginia, Charlottesville, VA 22904, USA \\ 
$^{2}$Institute for Advanced Study, Tsinghua University, Beijing, 100084, People's Republic of China \\
$^{3}$Harvard-Smithsonian Center for Astrophysics, 60 Garden Street, Cambridge, MA 02138, USA \\
$^{4}$National Astronomical Observatory of Japan, Mitaka, Tokyo 181-8588, Japan\\
$^{5}$Department of Astronomy, University of Illinois at Urbana-Champaign, Urbana, IL 61801, USA 
}

\date{Accepted XXX. Received YYY; in original form ZZZ}

\pubyear{2018}

\begin{document}
\label{firstpage}
\pagerange{\pageref{firstpage}--\pageref{lastpage}}
\maketitle

\begin{abstract}
Disk polarization in (sub)millimeter dust continuum is a rapidly growing field in the ALMA
era. It opens up the exciting possibility of detecting and characterizing magnetic fields
and grain growth in disks around young stellar objects. However, to use polarization for
probing the disk properties, its production mechanism must be ascertained first. To date,
the conventional mechanism involving magnetically aligned grains fails to explain the
polarization patterns detected in most disks. This is especially true for the inclined disk
of HL Tau in ALMA Band 3 (wavelength $\sim 3$~mm), which has an elliptical polarization pattern. The
elliptical pattern was taken as evidence for polarized emission by dust grains aligned with
their long axes perpendicular the direction of the radiative flux. We show that the
radiatively aligned grains produce a circular, rather than elliptical,
polarization pattern even in inclined disks such as HL Tau. An elliptical polarization
pattern can be produced if the grains are aligned aerodynamically by the difference in
rotation speed between the dust and gas through the Gold mechanism. However, a strong azimuthal variation in polarized
intensity is expected for both the radiative and aerodynamic alignment, but not observed in
the HL Tau disk in ALMA Band 3. We conclude that neither of these two mechanisms alone can
explain the data and the origin of the 3~mm polarization remains a mystery. We
speculate that this mystery may be resolved by a combination of both direct emission and
scattering by aerodynamically aligned grains. 

\end{abstract}

\section{Introduction}
The magnetic field is thought be one of the main drivers of the dynamics
and evolution of protoplanetary disks, through either magnetic-rotational 
instability (MRI; \citealt{Balbus1991}) or magnetic disk wind (\citealt{Blandford1982}; \citealt{Turner2014}). Obtaining firm observational evidence for the magnetic field is therefore 
one of the most sought-after goals of disk research. One of the most widely used methods for probing magnetic fields is through polarization of thermal dust emission, based on the theory of magnetic alignment of dust grains (\citealt{DG1951,Purcell1979,Dolginov1976,Hoang2018}; see \citealt{Lazarian2007} and \citealt{Andersson2015} for recent reviews). 
This method has been applied successfully to a wide range of scales, from molecular clouds ($\sim$pc or larger; e.g., \citealt{PlanckXIX2014,Fissel2016}) to protostellar envelopes ($\sim 100\sim 1000$ AU; e.g., \citealt{Girart2006,Stephens2013,Hull2014,Cox2018}). 

On the disk scale  ($\lesssim 100$ AU), evidence for the magnetic field has been difficult to obtain from the polarized dust emission. The first spatially resolved polarization in a T Tauri disk was detected 
in HL Tau through 1.3 mm observations from the Combined Array for Research in 
Millimeter-wave Astronomy (CARMA; \citealt{Stephens2014}). 
It shows a roughly uniform polarization pattern along the disk minor direction which, if interpreted as coming from magnetically aligned grains, would imply an uni-direction magnetic field along the major axis, which is unexpected for a rotating disk. Soon after the appearance of the theory of (sub)millimeter polarization through dust self-scattering \citep{Kataoka2015}, it became clear that the CARMA observed pattern in HL Tau is more consistent with scattering-induced polarization in an inclined disk \citep{Yang2016a,Kataoka2016a} than that produced by grains aligned by the widely expected toroidal magnetic fields, although grain alignment with more complex magnetic fields cannot be ruled out (\citealt{Stephens2014,Matsakos2016}; see also \citealt{Alves2018} for the case of BHB07-11). 
With polarimetric observations by the Atacama Large Millimeter/submillimeter Array (ALMA), spatially resolved polarization has been detected in an increasing number of circumstellar disks, especially at ALMA Band 7/6 
($0.87$ mm/$1.3$ mm). To date, the majority of the observed patterns are consistent with that from self-scattering, e.g., HL Tau (\citealt{Stephens2017}, Band 7),  IM Lup (\citealt{Hull2018} Band 7), IRS 63 (Sadovoy et al. in prep, Band 6), HH212 (\citealt{Lee2018}; Band 7), and HH80/81 (\citealt{Girart2018}, Band 6). 

Besides magnetically aligned grains and dust self-scattering, there are other mechanisms for producing millimeter/submillimeter polarization. One of such mechanisms, radiative alignment, was recently proposed by \cite{Tazaki2017}, based on the earlier work by \cite{LH2007}. 
Within the framework of grain alignment by radiative torque, radiative alignment could happen when the magnetic field is weak. In this case, grains will align with their long axes perpendicular to the local radiation flux, or the direction of the local radiation anisotropy, rather than the magnetic field. The mechanisms of radiative alignment and dust self-scattering have in common that they both rely on the anisotropy in the radiation field incident on the dust grain to produce polarization. Nevertheless, they have different dependence on wavelength and disk orientation, which makes them distinguishable, especially through multi-wavelength polarization observations and in disks with extreme inclinations (i.e., edge-on). 
%
%

\begin{figure}
\centering
\includegraphics[width=\columnwidth]{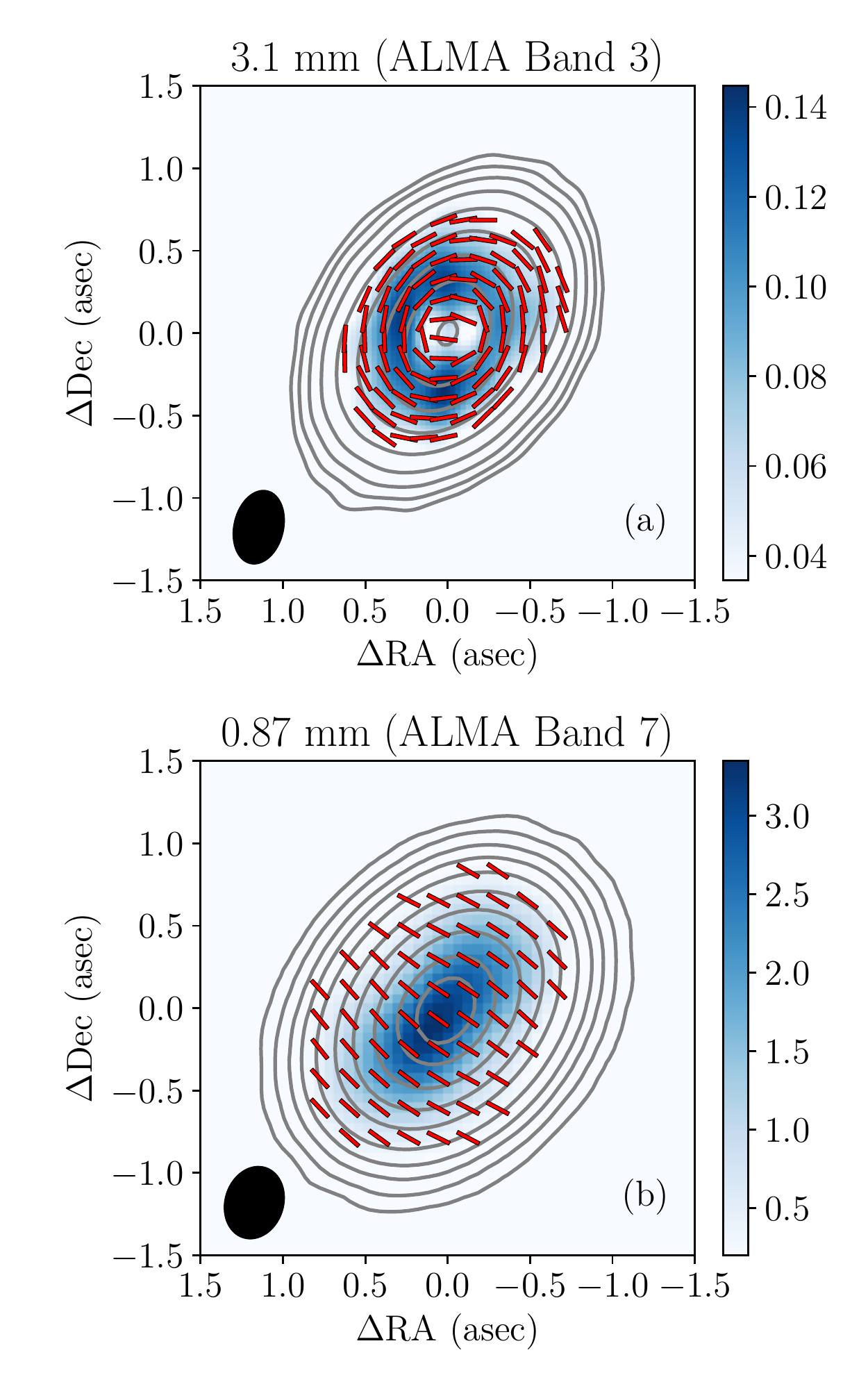}
\caption{HL Tau disk polarization detected by ALMA in Band 3 (panel a) and Band 7 (b). The panels are adopted from \protect\cite{Kataoka2017} and \protect\cite{Stephens2017}, respectively. Plotted are the polarization orientations (line segments with the same length independent of the  polarization fraction, E-vectors), polarized flux (color map) and total flux (contours). }
\label{fig:observation}
\end{figure}

To date, the strongest support for the mechanism of radiative alignment proposed by \cite{Tazaki2017} comes from the well-resolved polarization pattern detected in HL Tau by ALMA in Band 3 ($\sim 3$ mm; \citealt{Kataoka2017}; reproduced in Fig.~\ref{fig:observation}a). It has a broadly azimuthal pattern that is very different from the more or less uni-directional pattern detected in Band 7 ($\sim 0.87$~mm; \citealt{Stephens2017}; reproduced in Fig.~\ref{fig:observation}b); the latter is a  textbook example of what the scattering-induced polarization should look like in an inclined disk \citep{Yang2016a,Kataoka2016a}. Since the grains responsible for the Band 7 polarization would not produce any detectable polarization through scattering in Band 3 (because the scattering cross-section drops rapidly with wavelength in the Rayleigh limit), it is natural for \cite{Kataoka2017} to attribute the Band 3 polarization to radiative alignment rather than scattering; the conventional interpretation involving magnetically aligned grains would imply a magnetic field that is mostly radial in the disk plane, which is unlikely in a differentially rotating disk). More importantly, since the radiative flux in an axisymmetric disk is in the radial direction, the radiatively aligned grains are expected to have their long axes in the azimuthal direction in the disk plane, which is thought to produce a polarization pattern in the plane of the sky broadly resembling the observed pattern. 
%
%

In this paper, we show that there are two problems with the interpretation of the HL Tau Band 3 polarization coming from radiative alignment. For one, the Band 3 data do not follow exactly the polarization orientations expected from radiative alignment. The second, more severe problem is that the radiative alignment mechanism predicts a well-defined azimuthal variation in polarized intensity that is inconsistent with the data. 
In the rest of the paper, we will discuss these two problems 
quantitatively. 

The problem with polarization orientation is discussed in Sec.~\ref{sec:p1}, and that 
with azimuthal variation in polarized intensity is addressed in Sec.~\ref{sec:p2}. 
We find that the elliptical polarization orientation can be better explained by 
the \cite{Gold1952} mechanism of grain alignment (which we 
will refer to as ``aerodynamic alignment'' hereafter, since it relies
on the aerodynamic interaction between the gas and dust as they orbit
the central star at different speeds\footnote{We ignore the drift of the dust particles relative to the gas in the radial direction, which is typically slower than that in the azimuthal direction, as long as the Stokes number of the grains is not close to unity.}) 
than by radiative alignment and that both alignment mechanisms are expected to produce pronounced azimuthal variation in polarized intensity that should be easily observable. We compute the expected ALMA Band 3 polarization patterns for HL Tau disk based on the radiative and aerodynamic alignment mechanisms taking into account of the finite telescope beam, and compare them with the observational data in Sec.~\ref{sec:data}. We find that both mechanisms fail to match the polarization data and are thus disfavored. In Sec.~\ref{sec:discussion}, we stress the importance of taking into account beam-averaging in comparing model predictions and observational data and explore plausible ways to improve the model predictions. Additional challenges of aligned grain models in explaining the multi-wavelength observations of the HL Tau disk are briefly discussed. Our main results are summarized in Sec.~\ref{sec:summary}.

\section{Polarization orientation}
\label{sec:p1}
\subsection{Polarization pattern from radiative alignment in an inclined disk is circular, not elliptical}\label{sec:ori}

\begin{figure}
\centering
\includegraphics[width=\columnwidth]{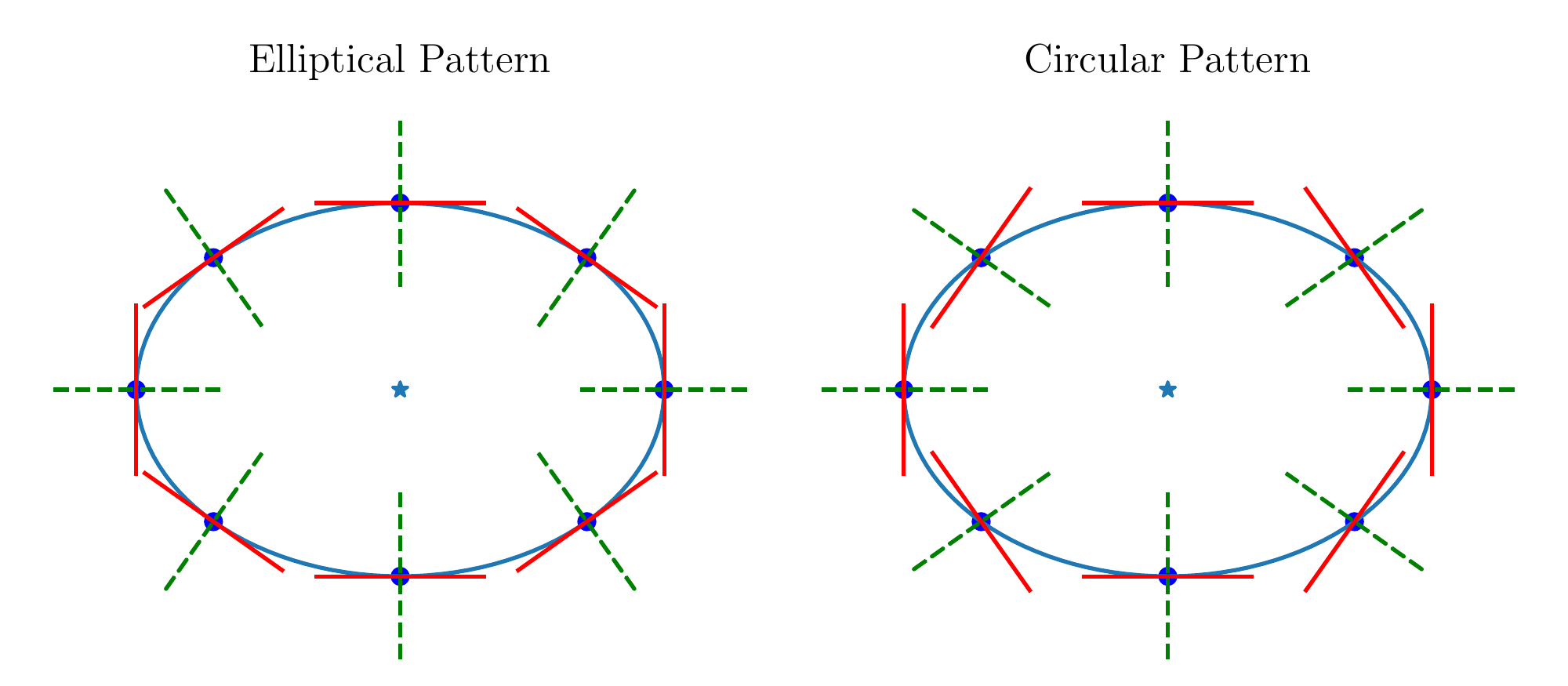}
\caption{Elliptical vs circular polarization pattern. The red solid line segments are for polarization orientations, and green dashed line segments for the direction of the required radiative flux projected into the sky plane in the case of radiative alignment.}
\label{fig:torvsrad}
\end{figure}

A major reason that the HL Tau polarization in ALMA Band 3 was attributed to radiative alignment was that the polarization vectors appear to follow the elliptical contours of constant brightness (which are the circles of constant dust emission in the disk plane projected onto the sky plane; see Fig.~\ref{fig:observation}a) and the radiative alignment was thought to produce such an elliptical pattern (see Fig.~3b of \citealt{Kataoka2017}). The expectation would be true if the grains have their shortest axes aligned by radiative flux along the radial direction in the disk plane {\it and} their longest axes staying in the disk plane (i.e., parallel to the local tangent of the circle in the disk plane that passes through the grain). In such a case, the long axes of the grains projected in the sky plane would still be aligned with the tangents to the circles projected to the sky plane (i.e., the ellipses), as illustrated by the solid line segments in the left panel of Fig.~\ref{fig:torvsrad}. An intuitive way to visualize the situation is to imagine the extreme case where needle-like grains are aligned along circles in the disk plane. When viewed at an inclination rather than face-on, the ``needles" would remain ``painted" on the circles, which now become the ellipses in the sky plane, producing an elliptical pattern. 

However, this is not what happens in the case of radiative alignment. Even though the shortest axes of the radiatively aligned grains are expected to be in the radial direction in the disk plane, their longest axes will not stay in the disk plane because of the spin around their shortest axes\footnote{Even in the absence of any spin, the longest axes would be distributed randomly in the plane perpendicular to the radial direction.}. The net effect is that the radiatively aligned grains are effectively oblate (due to spin or ensemble average), with their shortest axis (which is also the symmetry axis for the effectively oblate shape) in the radial direction. In this case, the short axis of the projected grain shape in the sky plane remains aligned with a radial line that passes through the center (see the green dashed line segments in the right panel of Fig.~\ref{fig:torvsrad} for illustration). To visualize the situation better, it is again helpful to go to the extreme case, where the effectively oblate grains are infinitely thin ``disks" (or ``flakes"). In this case, it is easy to show that, when projected to the sky plane, the ``disks" become ``ellipses" with their short axes along the radial direction in the sky plane and long axes perpendicular to the radial direction, producing a circular polarization pattern as illustrated by the red solid line segments in the right panel of Fig.~\ref{fig:torvsrad}. 
This is consistent with the well-known result of the polarization orientation from magnetically aligned grains in (optically thin) molecular clouds, which is always perpendicular to the B-field component in the sky plane (e.g., \citealt{Andersson2015}). If the elliptical pattern shown in the left panel of Fig.\ref{fig:observation} were to be produced by radiatively aligned grains (as previously envisioned, see \citealt{Kataoka2017}), the radiative flux would have to be oriented in such directions that, when projected into the sky plane, follow the green dashed lines, which would not go through the center (and thus not in the radial direction), contradicting the expectation in an axisymmetric disk.

\subsection{Aerodynamic alignment can produce elliptical polarization 
pattern}

Besides alignment by radiation field, grains can also be aligned aerodynamically when moving relative to the ambient gas \citep{Gold1952,Lazarian1995}. This is a possibility in circumstellar disks where the gas and dust orbit the central object at different speeds because the former experiences the gas pressure gradient directly but the latter does not. In the simplest case where the gas pressure increases radially inward, the gas would rotate at a sub-Keplerian speed because of the partial pressure support against the stellar gravity. Dust grains would rotate faster, and thus experiencing a ``head-wind"\footnote{It is also possible for the gas pressure to decrease radially inward, such as near the inner edge of a dense ring. In this case, the gas would rotate faster than the dust, again creating a relative motion between the two that is conducive to aerodynamic grain alignment.}. The relative speed between the gas and dust depends on several factors, particularly the gas density, grain sizes, 
and especially the relative speed between the dust and gas. In particular, efficient grain alignment through this mechanism may require a supersonic relative motion (e.g., \citealt{Hoang2014}), which may be difficult to achieve in a relatively quiescent protoplanetary disk.  Whether 
the mechanism can align the grains emitting in the ALMA Bands or not remains to be determined; we will postpone a detailed treatment of this mechanism to a future investigation. In what follows, we will argue that the polarization pattern expected from this mechanism is elliptical rather than circular, unlike the case of radiative alignment. 


The aerodynamically aligned grains are expected to have their longest axes along the ``streaming direction'', the direction of the relative movement between the gas and dust, which is in the azimuthal direction in the disk plane. If the grains precess rapidly around the streaming direction, they would have an ``effective" prolate shape. Even in the absence of any precession, ensemble-averaging of a large number of grains with their long axes preferentially aligned along the same (streaming) direction would also yield an ``effective" prolate shape. As discussed earlier in \S~\ref{sec:ori}, prolate grains with their long axes aligned along the azimuthal direction in the disk plane produce an elliptical rather than circular polarization pattern in the sky plane. Again, this can be visualized most easily in the extreme case of ``needle-like" grains.   
%
%

\subsection{Differences between circular and elliptical patterns}

In this subsection, we quantify the expected difference in polarization orientation between the elliptical and circular patterns illustrated in Fig.~\ref{fig:torvsrad} under the assumption of a optically and geometrically thin (dust) disk and discuss whether such a difference is measurable in the HL Tau ALMA Band 3 data. 



For the circular pattern, the polarization angle $\theta_\mathrm{cir}$ at any point in the sky plane is simply the azimuthal angle of that point in the sky plane $\theta_\mathrm{sky}$ rotated by $90^\circ$, i.e. the polarization is always perpendicular to the radial direction in the sky plane, namely 
\begin{equation}
\theta_\mathrm{cir} = \theta_\mathrm{sky} + \frac{\pi}{2}.
\label{eq:t_cir}
\end{equation}

For the elliptical pattern, the polarization angle $\theta_\mathrm{ell}$ at a given point depends on the shape of the ellipse, which in turn is controlled by the inclination of the disk $i$ ($i=0$ for face-on view). It is related to the azimuthal angle $\theta_\mathrm{sky}$ of that point (measured relative to the major axis of the projected disk or ellipse)
through 
\begin{equation}
\tan(\theta_\mathrm{ell}) = -\frac{\cos^2(i)}{\tan(\theta_\mathrm{sky})}. 
\label{eq:t_ell}
\end{equation}

\begin{figure}
\centering
\includegraphics[width=\columnwidth]{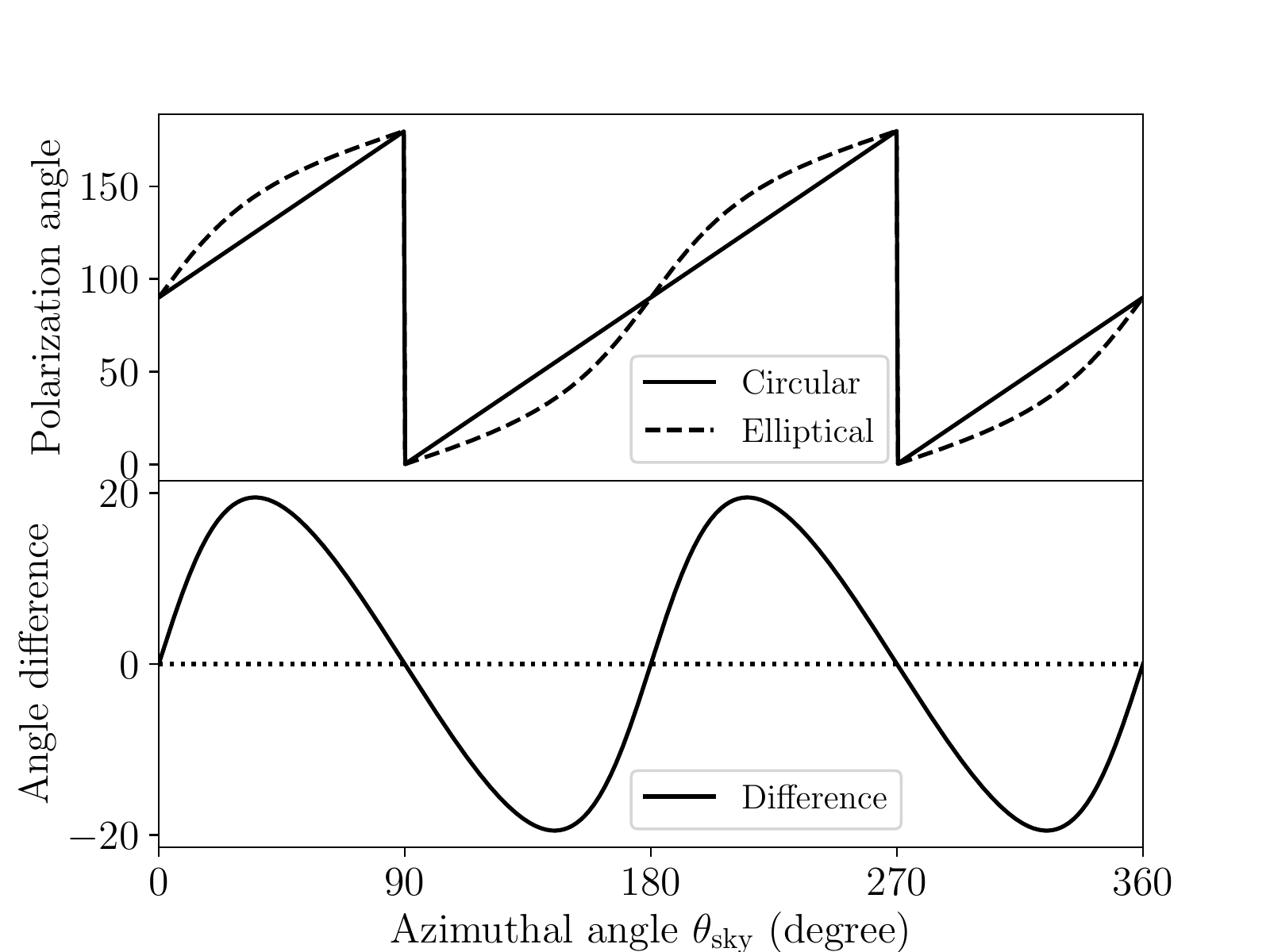}
\caption{Top panel: expected polarization angle for circular ($\theta_\mathrm{cir}$; solid line) and elliptical ($\theta_\mathrm{ell}$; dashed) pattern as a function of the azimuthal angle $\theta_\mathrm{sky}$ in the sky plane for a disk inclined by $45^\circ$ to the line of sight. 
Bottom panel: the difference in polarization orientations between the two patterns, which can be as large as $\sim 20^\circ$ for this disk inclination.
}
\label{fig:angles}
\end{figure}

Fig. \ref{fig:angles} shows the polarization angle, which ranges from $0^{o}$ to $180^{o}$, as a
function of azimuthal angle in the sky plane, for the case of a $45^{o}$ 
inclined disk (similar to HL Tau disk). The top panel shows the expected orientation for both 
circular pattern (solid line) and elliptical pattern (dashed line). The
bottom panel shows the difference in polarization angle between two patterns. We can see that
the difference can be as large as $\sim 20^\circ$. This difference should be distinguishable with the current data over most of the region of interest. For example, near the peak of the 
polarized  intensity at the HL Tau Band 3, the signal-to-noise ratio was 
reported to be $21~\sigma$ \citep{Kataoka2017}, which roughly corresponds to an error of $2^\circ$ in polarization angle.


\begin{figure}
\centering
\includegraphics[width=\columnwidth]{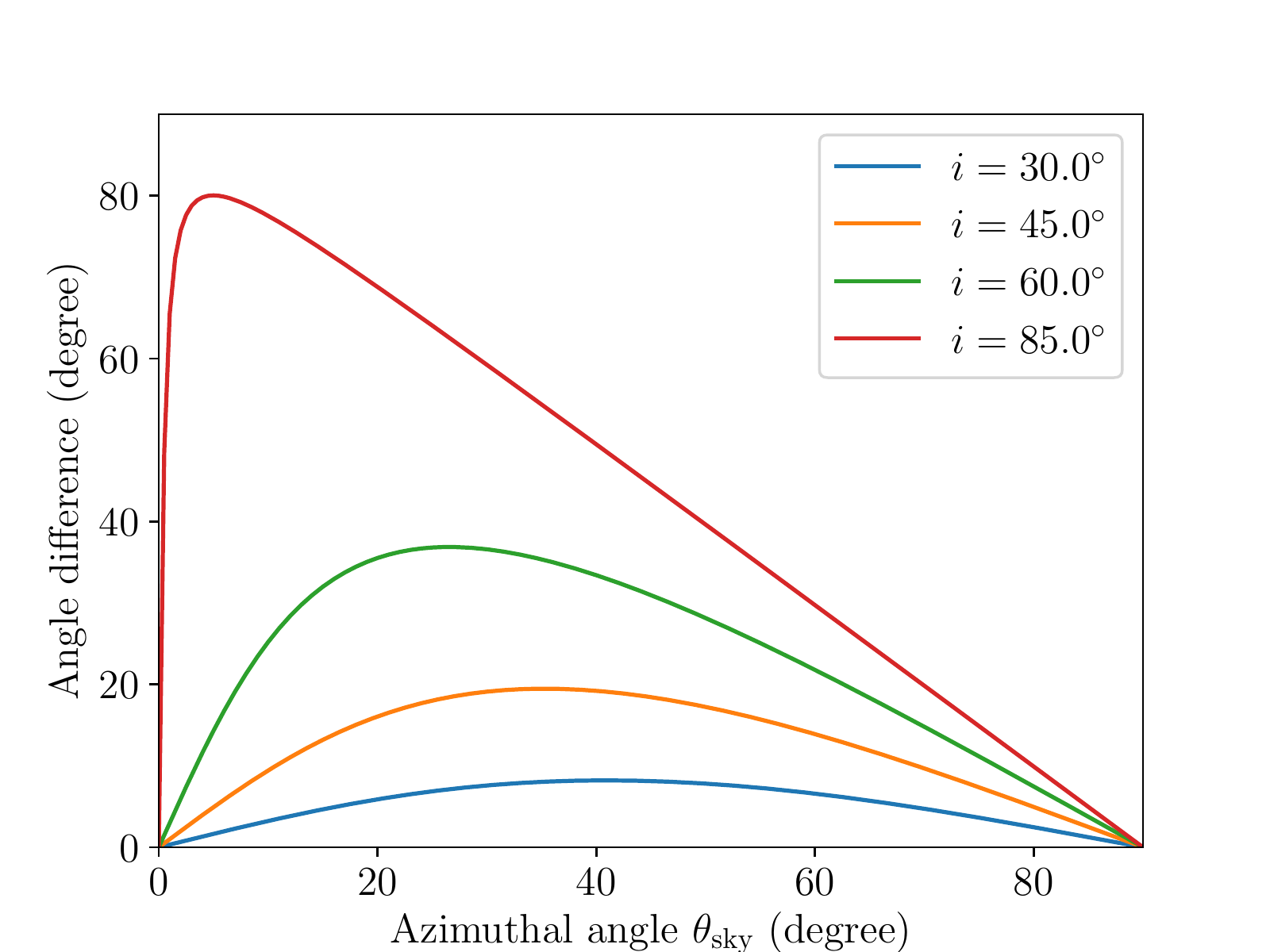}
\caption{The difference in polarization angle between the circular and elliptical patterns as a function of the azimuthal angle in the sky plane 
$\theta_\mathrm{sky}$ between $0^\circ$ (major axis) and $90^\circ$ (minor axis), for disks inclined by different angles ($i=0^\circ$ for face-on view). The difference is larger for a more inclined disk. 
%
%
}
\label{fig:angles_inc}
\end{figure}

\begin{figure}
    \centering
    \includegraphics[width=\columnwidth]{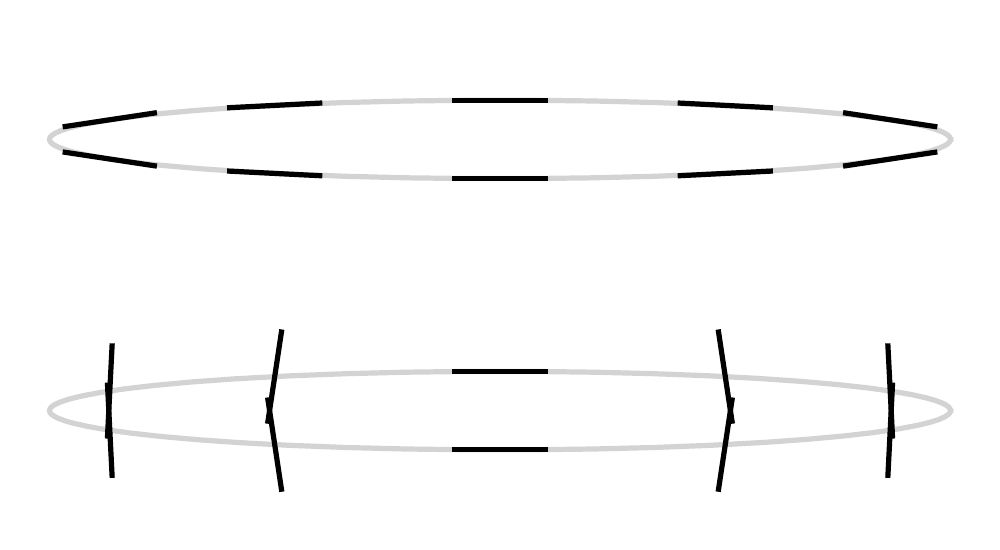}
    \caption{Large difference in polarization orientation between the elliptical (upper panel) and circular (lower) pattern for a nearly edge-on disk (with an inclination angle $i=85^\circ$). Such disks are ideal for distinguishing the two patterns. 
    }
    \label{fig:85v1}
\end{figure}

Fig. \ref{fig:angles_inc} shows the difference in polarization angle between the two polarization patterns for different disk inclination angles. These two patterns are the same for face-on disks ($i=0^\circ$), as expected. Their difference increases as the disk becomes more inclined to the line of sight, reaching $90^\circ$ near the major axis (where the azimuthal angle measured from the major axis, $\theta_\mathrm{sky}$, approaches $0^\circ$) for an edge-on disk. For a given inclination angle $i$ (that is not exactly $90^\circ$), the difference vanishes on the major ($\theta_\mathrm{sky}=0^\circ$) and minor ($\theta_\mathrm{sky}=0^\circ$) axis, and peaks at a location in the sky plane that is closer to the major axis than the minor axis. 
%
Fig.~\ref{fig:85v1} illustrates pictorially the nearly edge-on case of $i=85^\circ$, where the polarization orientations in the elliptical and circular patterns are almost orthogonal over most of the (narrow, projected) disk. It is in such cases that the difference between the two patterns is most easily distinguishable\footnote{Edge-on disks are often optically thick, which could complicate the interpretation of the observed polarization pattern (e.g., \protect\citealt{Yang2017}). }

%

The difference in polarization orientation between the elliptical and circular patterns translates to a difference in polarization orientation in unresolved disks. This difference is best illustrated by the nearly edge-on case (see Fig.~\ref{fig:85v1}), where most of the polarization vectors are roughly parallel to the major axis (or the narrow, projected disk) for the elliptical pattern (yielding an averaged polarization along the major axis), but largely perpendicular to it for the circular pattern (yielding an averaged polarization along the minor axis). 
This difference persists for more moderately inclined disks as well. 

To illustrate the difference in averaged polarization more quantitatively, we will consider the simplest case where the polarization intensity and fraction are spatially uniform across the disk in the sky plane. The disk-averaged polarization fraction for the circular pattern becomes 
\begin{equation}
{\bar p}_{\rm cir} = -p_0 \frac{1-\mathrm{cos}(i)}{1+\mathrm{cos}(i)}
\label{eq:po_cir}
\end{equation}
where the subscript ``cir" denotes ``circular pattern" rather than ``circular polarization," $p_0$ is the polarization fraction at each point before average, $i$ is the disk inclination angle ($i=0^\circ$ for face-on), and ${\bar p} < 0$ means polarization along minor axis of the projected disk.  Similarly, the disk-averaged polarization fraction for the
elliptical pattern is: 
\begin{equation}
{\bar p}_{\rm ell} = p_0 \frac{1-\mathrm{cos}(i)}{1+\mathrm{cos}(i)}
\label{eq:po_ell}
\end{equation}
which is positive, and thus along the major (rather than the minor) axis.  The opposite sign in the averaged polarization fraction is a generic difference that can in principle be used to distinguish the two patterns. In this particular example, the magnitude of the averaged polarization fraction is the same for the two patterns. This is not true in general, especially when the expected azimuthal variation of the polarization fraction is taken into account (see section \ref{sec:p2} below). 

Nonetheless, Eqs.~(\ref{eq:po_cir}) and (\ref{eq:po_ell}) reveal an interesting point. Even though the elliptical and circular polarization patterns have a high degree of symmetry, their disk-averaged polarization fraction ${\bar p}$ is reduced from the intrinsic value $p_0$ but does not vanish completely in an inclined disk. The reduction factor depends on the disk inclination angle $i$ and the spatial distribution of the polarization intensity. In the simplest case of spatially constant polarization considered above, we have ${\bar p}=(\sqrt{2}-1)/(\sqrt{2}+1) p_0\approx 0.172\,p_0$ for $i=45^\circ$. It increases with the inclination angle, approaching the intrinsic value $p_0$ in the limit of an edge-on disk.

\section{Azimuthal variation of polarization degree}
\label{sec:p2}

Besides the polarization orientation, the spatial variation of polarization fraction, especially in the azimuthal direction, is also an important discriminant between different polarization mechanisms. In this section, we will concentrate on the azimuthal variation of the polarization fraction expected from radiative alignment, and contrast it with those from other mechanisms, especially aerodynamic alignment. 

\subsection{Polarization dependence on the inclination of grain alignment axis to the line of sight}

The polarization of the thermal emission from aligned non-spherical dust grains depends on the ellipticity of the grains as viewed by the observer in the sky plane. For grains that are effectively ``oblate'' (or ``disk-'' or ``flake-like" in the extreme case), e.g., when the alignment axis is the shortest axis of the grain (as true for magnetic and radiative alignment), the polarization is maximized when the ``disk'' is viewed edge-on, with its shortest (alignment) axis in the sky plane. We will denote this maximum polarization fraction by $p_0$, and refer to it as the ``intrinsic polarization." When the shortest (alignment, symmetry) axis of the effectively oblate grain is inclined by an angle $i_d$ to the line of sight, the polarization fraction becomes 
\begin{equation}
p(i_d) = \frac{p_0\ \mathrm{sin}^2(i_d)}{1+p_0\ \mathrm{cos}^2(i_d)}
\label{eq:pinc_o}
\end{equation}
in the dipole regime appropriate for small grains \citep{LeeDraine1985,Yang2016b}. 
Note that $p(i_d=\pi/2) = p_0$, which recovers the intrinsic polarization fraction for grains viewed edge-on.  Since the maximum polarization is observed to be of order $1-10\%$ on the disk scale, we have, to a good approximation, $p(i_d)\approx p_0\ \mathrm{sin}^2(i_d)$.

For grains that are effectively ``prolate'' (or  ``needle-like'' in the extreme case), e.g., when the alignment axis is the longest axis, as true for aerodynamic alignment, the polarization is maximized when the longest (alignment) axis of the ``needle'' is in the sky plane. We again denote this maximum or intrinsic polarization fraction by $p_0$. In the more general case, we have 
\begin{equation}
p(i_d) = \frac{p_0\ \mathrm{sin}^2(i_d)}{1-p_0\ \mathrm{cos}^2(i_d)}
\label{eq:pinc_p}
\end{equation}
where $i_d$ is the inclination angle of the longest (alignment, symmetry) axis of the grain to the line of sight. The equation is again derived under the dipole approximation \citep{LeeDraine1985}. 

\subsection{Azimuthal variation of polarization degree in inclined disks}

\begin{figure}
    \centering
    \includegraphics[width=\columnwidth]{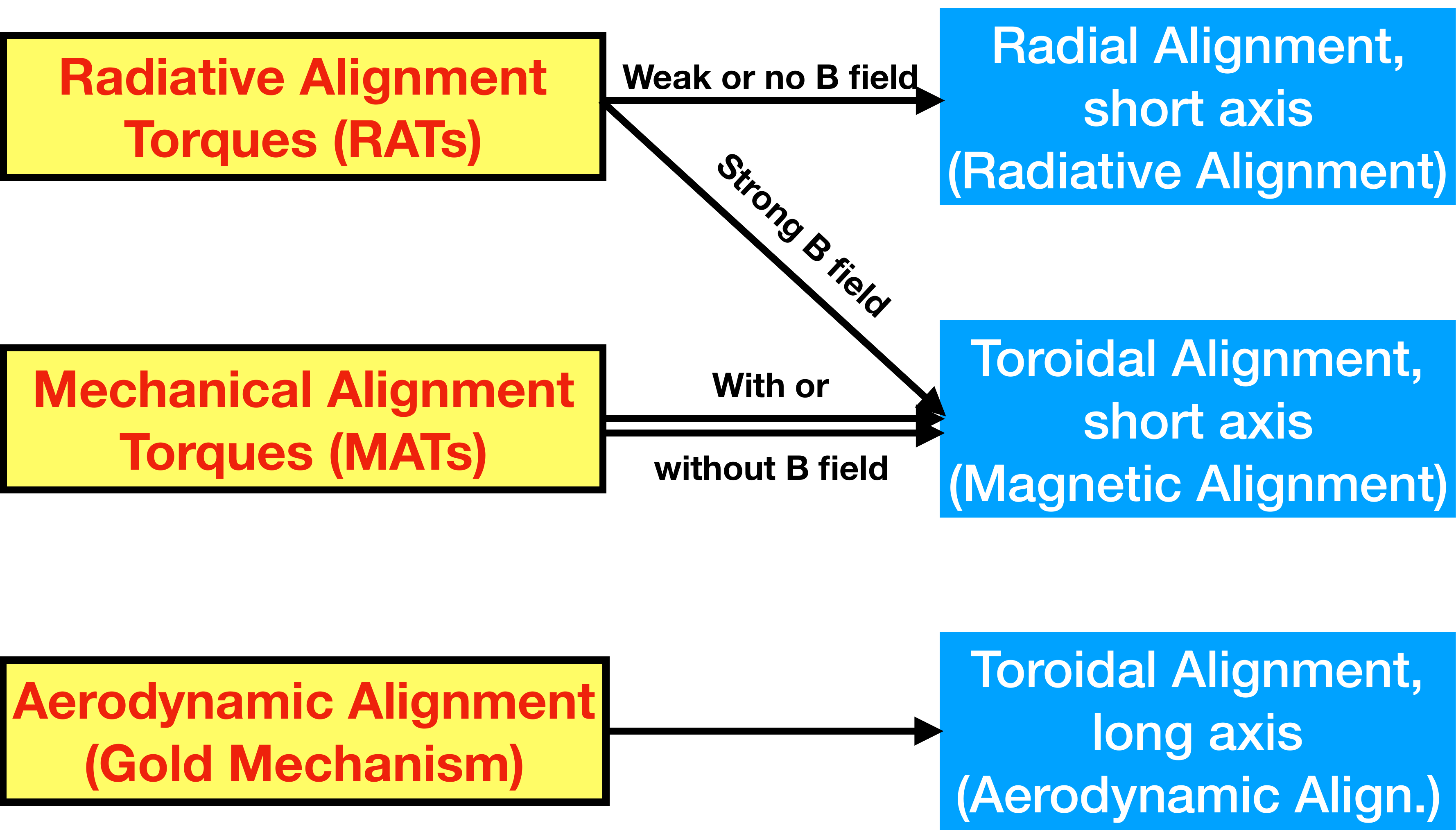}
    \caption{The most simplistic and final alignment state of the
    dust grains for various alignment mechanisms. See text for more 
    information.
    }
    \label{fig:newChart}
\end{figure}

\begin{figure}
\centering
\includegraphics[width=\columnwidth]{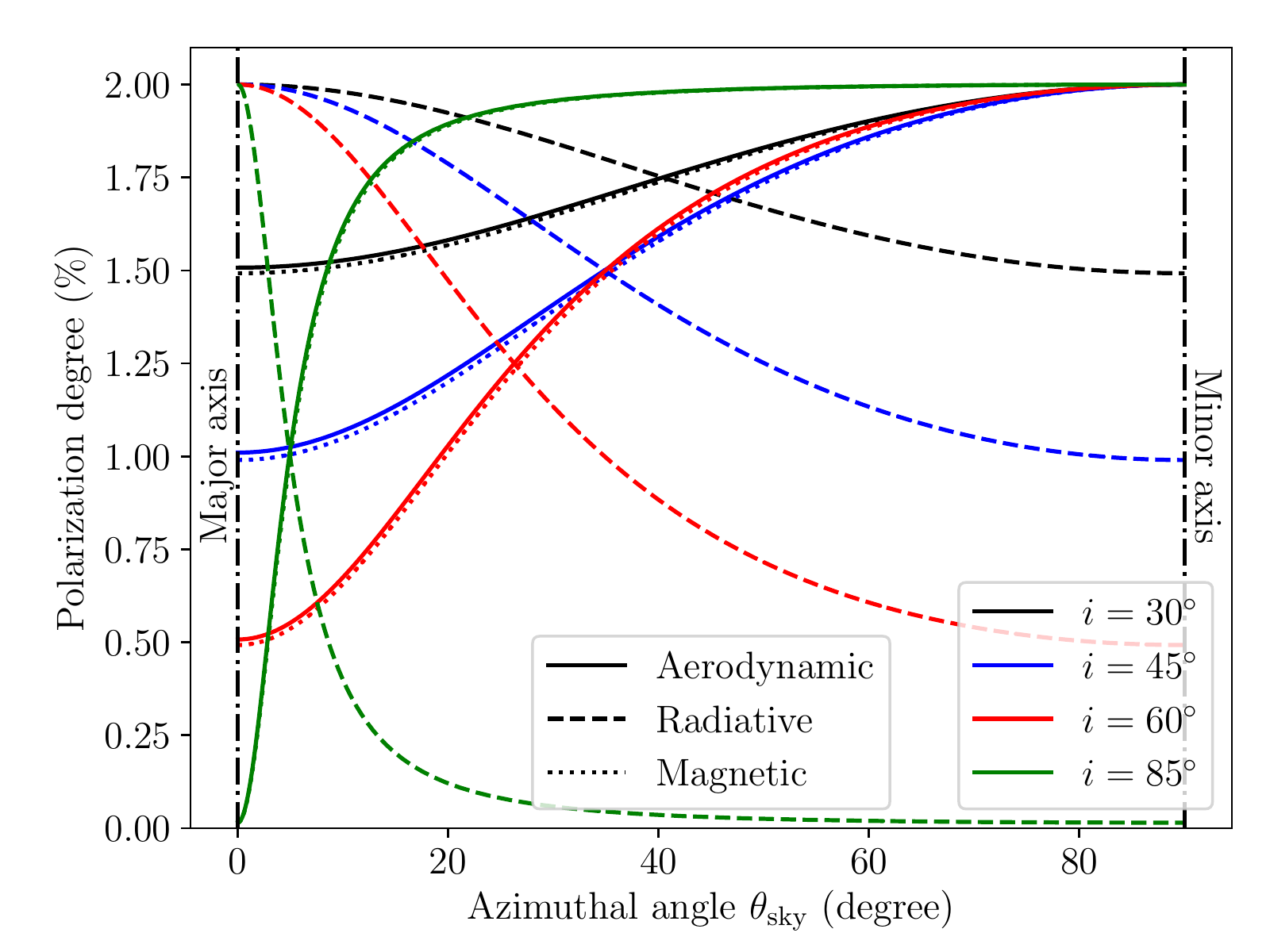}
\caption{Polarization degree as a function of the azimuthal angle in the sky plane $\theta_\mathrm{sky}$ for different alignment mechanisms and different
inclination angles. The polarization degrees from the magnetic alignment (dotted line) and aerodynamic alignment (solid) have a similar angular dependence, both peaking on the minor axis ($\theta_\mathrm{sky}=90^\circ$). This is opposite to that of the radiative alignment (dashed), which peaks on the major axis ($\theta_\mathrm{sky}=0^\circ$) instead. 
Different inclination angles are represented by different colors. Note that the polarization degree on the minor axis decreases monotonically with increasing inclination angle for the case of radiative alignment (dashed lines), vanishing completely in the edge-on case with $i=90^\circ$. 
}
\label{fig:pang}
\end{figure}

Given the above analytical expressions, we can now quantify the azimuthal variation of the polarization degree for aligned grains in inclined disks. We will consider different alignment mechanisms, including  magnetic, radiative, and aerodynamic alignment. 
Note that we only consider the most simplistic and final alignment state of the dust grains, which are expected to fall into one of these three scenarios 
(see Fig.~\ref{fig:newChart} for a chart summary).
RATs predict alignment with a magnetic field in the presence of a strong magnetic field, which is the toroidal magnetic alignment considered 
here.\footnote{We should note that a disk magnetic field of tens  of mG in strength may not be strong enough to align large millimeter-sized grains efficiently unless the number of iron atoms per ferromagnetic cluster envisioned in \cite{JonesSpitzer1967}, i.e., the quantity $N_{cl}$ in equation (1) of \cite{Hoang2017}), is large enough in the grains (see also \citealt{HoangLazarian2016}). }
In the absence of a magnetic field, RATs produce grains aligned with radial radiation flux, which is the radiative alignment here. The Gold mechanism produces the alignment of the long axes of the grains in the azimuthal direction, which is the aerodynamic alignment case considered here. A newly developed Mechanical Alignment Torques theory (MATs)\citep{LazarianHoang2017M, Hoang2018} also predicts magnetic alignment in the presence of a strong magnetic field. In the absence of a magnetic field, MATs may cause dust grains to align with short axis along the drifting direction between gas and grains. Here we assumed purely a toroidal drifting velocity and thus the alignment axis in this case is the same as a toroidal magnetic alignment. More complicated flow directions will indeed generate more complex polarization features (e.g., Kataoka et al. 2018, submitted). It is also plausible that multiple mechanisms can act together to produce polarization patterns that are significantly different from the three basic patterns considered in this paper, although special conditions may be required for this to happen over most of the disk.
The results are shown in Fig.~\ref{fig:pang}, which plots the polarization degree as a function of the azimuthal angle in the sky plane measured from the major axis for all three cases, assuming a maximum or intrinsic polarization fraction of $p_0=2\%$ for illustration purposes.   

For the magnetic alignment case shown in Figure~\ref{fig:pang} (dotted line), we assume a pure toroidal magnetic field. As discussed earlier, grains with their shortest axes aligned with the magnetic field are effectively oblate, or ``disk-like'' in the extreme case. In the simplest case of a face-on circumstellar disk, the effectively oblate grains are viewed ``edge-on" everywhere, with the grain alignment (symmetry) axis perpendicular to the line of sight, yielding the maximum polarization. In an inclined disk, the grains on the minor axis remain ``edge-on" to the line of sight, but those on the major axis are viewed by the observer more ``face-on" and thus rounder, yielding a lower polarization degree. Indeed, the polarization degree on the major axis is simply given by equation~(\ref{eq:pinc_o}) with the angle $i_d$ between the grain alignment axis and the line of sight given by $i_d=90^\circ-i$, where $i$ is the disk inclination angle. For the representative case of $i=45^\circ$ shown in Fig.~\ref{fig:pang} (blue dotted curve), we have $p=0.91\%$ (for the adopted $p_0=2\%$) on the major axis. The polarization degrees at intermediate angles between the minor and major axes can be computed using simple geometry (see also \citealt{Cho2007}, \citealt{Yang2016b}, \citealt{Bertrang2017}).  

For the radiative alignment case shown in Figure~\ref{fig:pang} (dashed lines), we make the usual assumption that the radiation flux is in the radial direction in the disk plane. Using the same argument as above, it is easy to show that the effectively oblate grains on the major axis remain ``edge-on" to the line of sight in an inclined disk (and thus emit maximally polarized light), while those on the minor axis are viewed more  ``face-on" and thus appear rounder to the observer, yielding a lower polarization degree, given by equation~(\ref{eq:pinc_p}) with $i_d=90^\circ-i$. 

For the aerodynamic alignment case shown in the figure (solid lines), we assume that the grains have their longest axes aligned with the azimuthal direction in the disk plane. As discussed earlier, such grains are effectively prolate or ``needle-like." On the minor axis of an inclined disk, the aligned grains always have their longest (alignment, symmetry) axes in the sky plane, yielding maximum polarization. Those on the major axis, on the other hand, have their longest (symmetry) axes inclined by an angle $i_d=90^\circ -i$ (where $i$ is the disk inclination angle) to the line of sight, and thus appear less elongated (i.e., rounder) to the observer, yielding a lower polarization, given by equation~(\ref{eq:pinc_p}). For the representative case of $i=45^\circ$ shown in  Fig.~\ref{fig:pang} (blue solid line), the polarization degree on the major axis is 
$p=1.1\%$ (for the adopted $p_0=2\%$), which is comparable to, but slightly larger than, that on the major axis for the magnetically aligned grains. Mathematically, the difference comes from the fact that the minus sign in the denominator of equation~(\ref{eq:pinc_p}) is replaced by a plus sign in equation~(\ref{eq:pinc_o}). Physically, it is due to the difference in the effective shape of the aligned grains (oblate for magnetic alignment vs prolate for aerodynamic alignment). Nevertheless, the azimuthal variations of the polarization degree for these two cases are very similar, both decreasing monotonically from the minor axis to the major axis. This trend is opposite to the case of radiative alignment, where the alignment axis is along the radial direction in the disk plane, rather than the azimuthal direction (as in the other two cases). 

\begin{figure}
    \centering
    \includegraphics[width=\columnwidth]{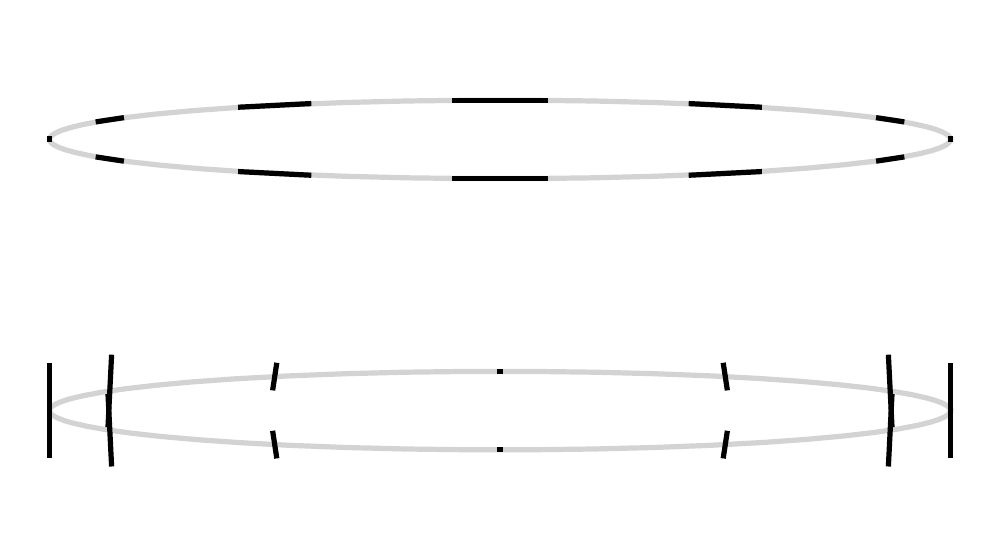}
    \caption{Large difference in not only polarization orientation but also azimuthal variation of polarization degree between the aerodynamic (upper panel) and radiative (lower) alignment for a nearly edge-on disk (with an inclination angle $i=85^\circ$). Such disks are ideally suited for distinguishing the two alignment mechanisms. }
    \label{fig:85v2}
\end{figure}

The difference in azimuthal variation of the polarization degree between the case of radiative alignment and those of magnetic and aerodynamic alignment increases with the inclination angle. To illustrate this difference more pictorially, we plot in Fig.~\ref{fig:85v2} the polarization pattern in a nearly edge-on disk with $i=85^\circ$, with the polarization degree proportional to the length of the line segments. This figure drives home the point that edge-on disks are ideal for distinguishing the different polarization mechanisms, from not only the polarization orientation (see also Fig.~\ref{fig:85v1} above) but also the azimuthal variation in polarization degree. 

 Note, in particular, the low polarization fraction near the disk center for the radiative 
alignment case in the nearly edge-on disk. The physical reason for this interesting behavior is that the grains near the center are aligned with radiative flux that is close to the line of sight, which makes the effectively oblate grains appear almost circular in the sky plane, yielding little polarization. This robust feature is especially useful for distinguishing the radiative alignment from other mechanisms, as discussed in \cite{Lee2018} and Harris et al. (in press). 

\section{HL Tau Band 3 polarization}
\label{sec:data}

\subsection{The model}
\label{sec:model}

\begin{figure*}
    \centering
    \includegraphics[width=\textwidth]{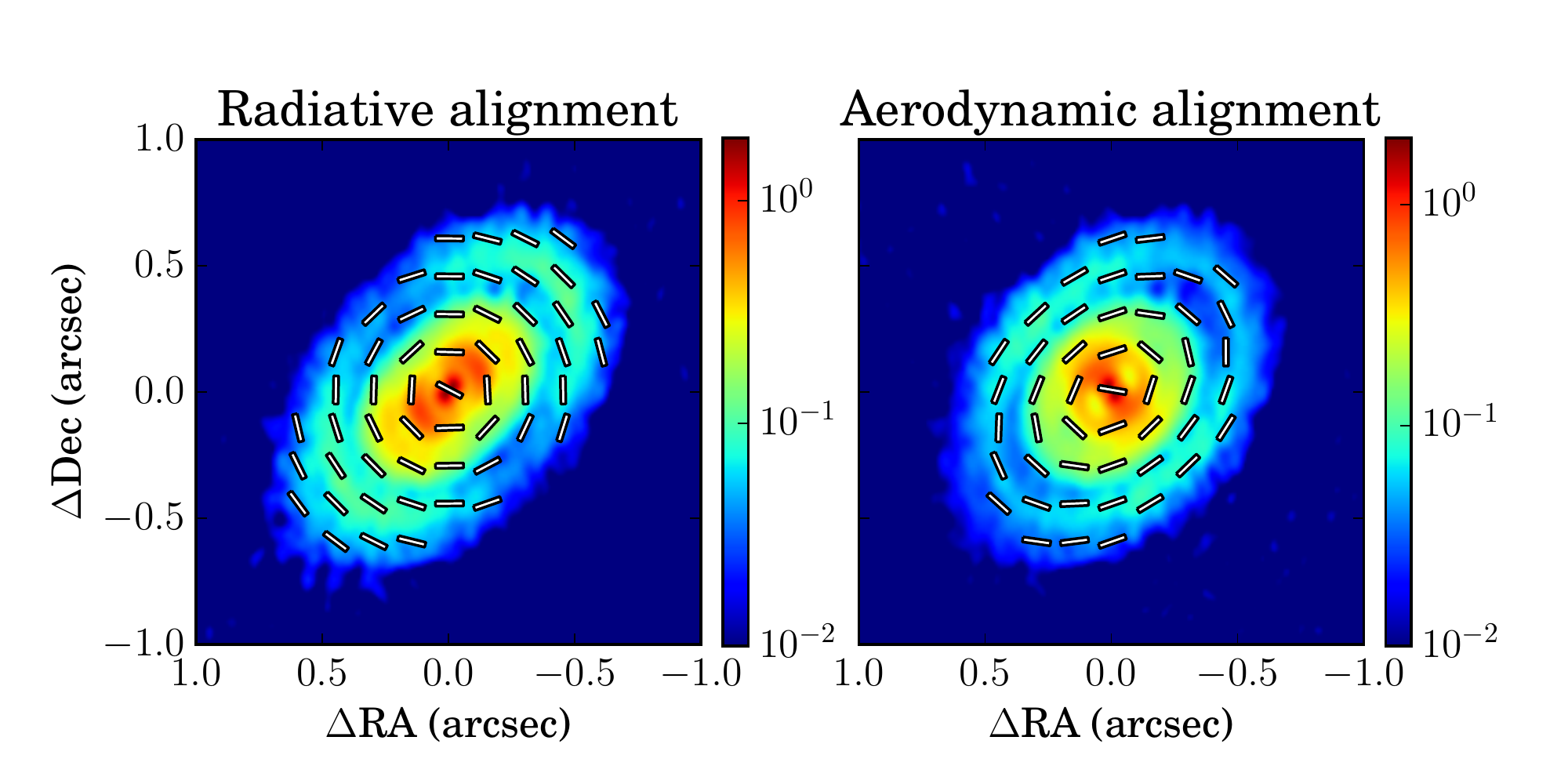}
    \caption{The modeled polarization without telescope beam convolution. The color map represents the polarized intensity (with arbitrary units). The vectors with uniform length show the orientation of the polarization. The differences between these two mechanisms are clear. Radiative alignment produces a circular polarization pattern with a stronger polarization along the major axis. Aerodynamic alignment produces an elliptical polarization pattern with a stronger polarization along the minor axis. }    
    \label{fig:noConv}
\end{figure*}

Of the three grain alignment mechanisms discussed in the last section, magnetic alignment is the least likely possibility for producing the Band 3 polarization observed in HL Tau disk (and shown in Fig.~\ref{fig:observation}a) because it predicts a polarization orientation perpendicular, rather than parallel, to the elliptical contours of iso-intensity. For the radiative alignment, we have already pointed out one of its potential problems: it predicts a circular, rather than elliptical (Fig.~\ref{fig:torvsrad}), polarization pattern that appears different from the observed pattern. The aerodynamic alignment mechanism may do better at matching the observed polarization pattern, but it predicts an azimuthal variation of the polarization degree (see Fig.~\ref{fig:pang}) that is not obvious in the data. The same problem is expected for the case of radiative alignment. In this section, we quantify the differences between the data and the model predictions based on the radiative and aerodynamic alignment, taking into account of the finite resolution of the ALMA observations, which is important for properly comparing the data and models because of beam smearing of both the polarization orientation and intensity distribution.  
%
%
%
%


For the polarization orientation, we adopt the circular pattern for the radiative alignment model and the elliptical pattern for the aerodynamic alignment model. For the radial variation of the polarized intensity, we use the much higher resolution ($0_{\cdot}^{\prime\prime}0853\times 0_\cdot^{\prime\prime}0611$) ALMA Band 3 continuum data from \cite{ALMA2015} as the Stokes I model, and assume a maximum (or intrinsic) polarization degree of $p_0=2\%$, comparable to the maximum observed value. The azimuthal variation of the polarization degree is then computed based on equations~(\ref{eq:pinc_o}) and (\ref{eq:pinc_p}) for a disk inclination angle of $i=46.72^\circ$ \citep{ALMA2015}, as done in the last section. 
The results without telescope beam convolution (discussed in more detail below in Sec.~\ref{sec:neither}) are shown in Fig.~\ref{fig:noConv}. We can clearly see the differences between these two models. On the one hand, the polarization from radiatively aligned grains forms a circular pattern, whereas the aerodynamic alignment produces an elliptical pattern. These two orientations are the same along the major and minor axes, and the difference becomes bigger at intermediate azimuthal angles. On the other hand, polarized intensity is concentrated along the major axis for radiative alignment, which is the opposite to that for aerodynamic alignment.

\subsection{Neither radiative nor aerodynamic alignment}
\label{sec:neither}
\begin{figure}
\centering
\includegraphics[width=\columnwidth]{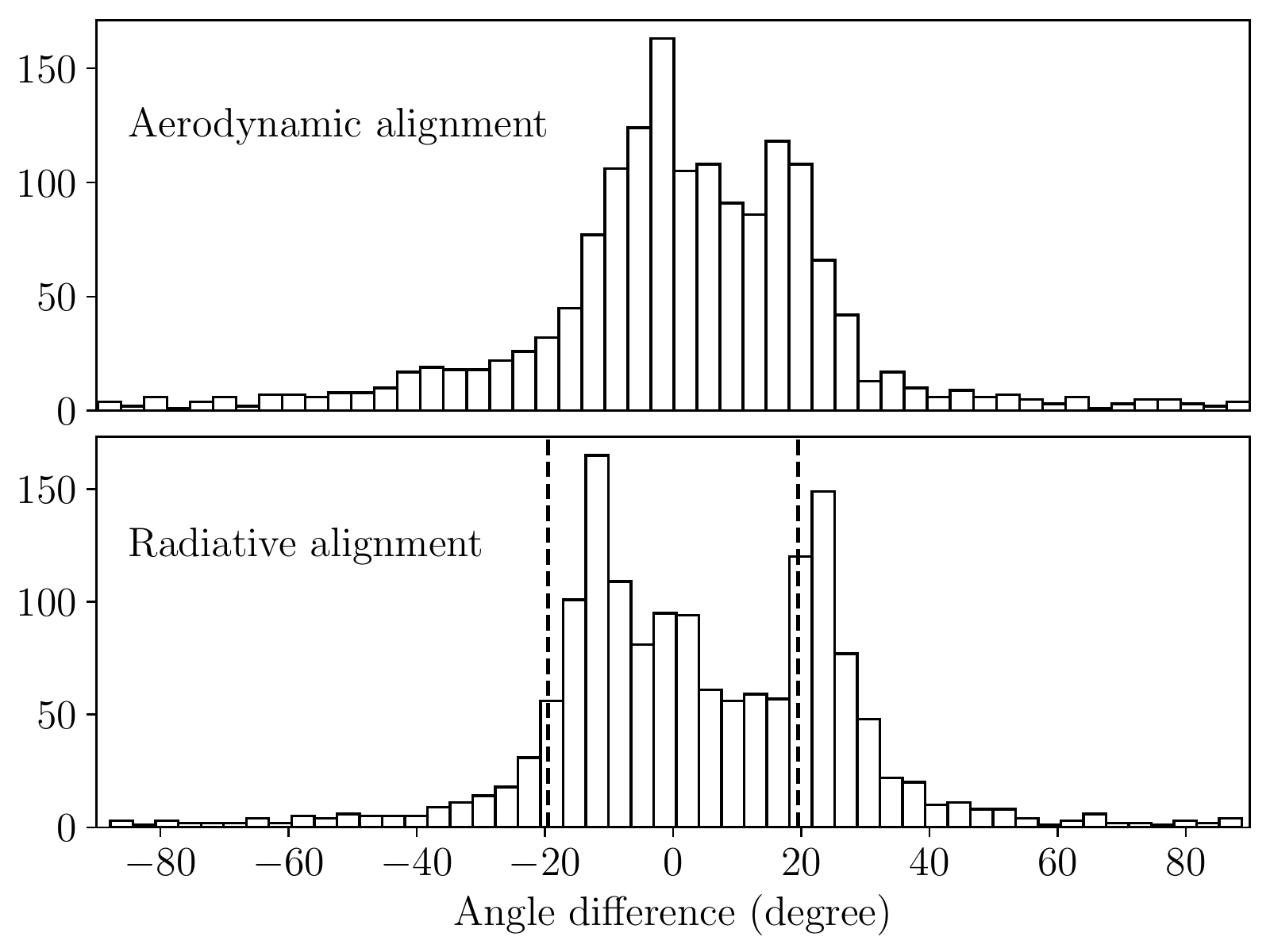}
\caption{Histogram of the difference in polarization orientation between the ALMA Band 3
data and the models based on the aerodynamic (upper panel) and radiative (lower) alignment. The two dashed vertical lines are at $\pm 19.5^\circ$, the maximum difference expected between the circular and elliptical polarization pattern for an inclined disk of $i=45^\circ$, as shown in Fig.~\ref{fig:angles}. }
\label{fig:hist}
\end{figure}

\begin{figure*}
\centering
\includegraphics[width=\textwidth]{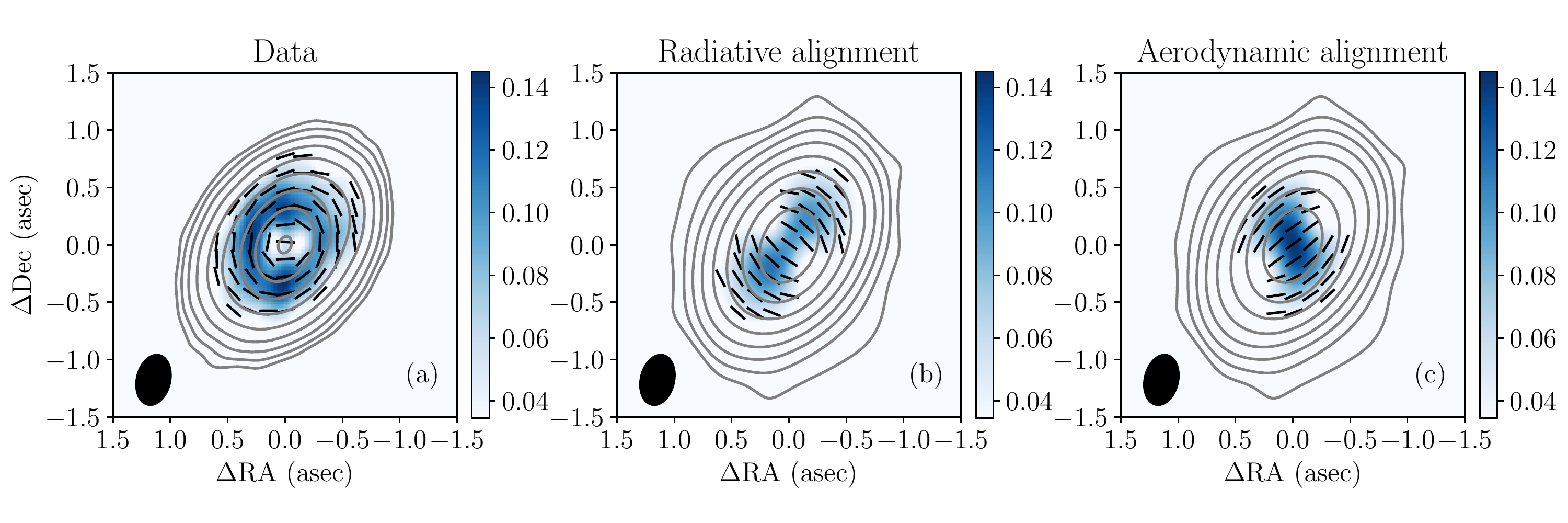}
\caption{Comparison between the ALMA Band 3 data (left panel) and the models based on radiative (middle) and aerodynamic (right) alignment. The colormap shows the polarized flux, in mJy/beam, and the contours are for Stokes I,
corresponding to (10, 20, 40, 80, 160, 320, 640, 1280, 2560)$\times$ the rms
of $9.6\rm\,\mu Jy$. 
The line segments denote the polarization orientations. They are plotted only in regions with polarized intensity above $34.5\rm\, \mu Jy/beam$, which corresponds to $5\sigma$ noise level in \protect\cite{Kataoka2017}. The ellipse at the lower corner represents the interferometric beam.}
\label{fig:maps}
\end{figure*}

The models in Fig.~\ref{fig:noConv}, generated from the ALMA Band 3 long-baseline observation of unpolarized continuum emission of \cite{ALMA2015}, have a much higher spatial resolution than Band 3 polarization observation by \cite{Kataoka2017}. To compare with real data, a telescope beam convolution is needed. We convolve the modeled Stokes I, Q, U map separately with the beam used for the Band 3 polarization observation ($0_\cdot^{\prime\prime}445\times 0_\cdot^{\prime\prime}294$). The results are shown in Figs. \ref{fig:hist} and \ref{fig:maps}. 
%
%

Fig.~\ref{fig:hist} shows the difference in polarization orientation between the Band 3 data and the model predictions at all locations on the disk where the polarization is detected at at least $5\sigma$ level. As expected, the aerodynamic alignment model reproduces the observed polarization orientations better, with the distribution of the angle difference centered around $0^\circ$. The relatively large dispersion around $0^\circ$ comes from beam smearing coupled with significant azimuthal variation of the polarized intensity (see discussion in the last section and Fig.~\ref{fig:maps} below). In contrast, the angle difference between the data and the radiative alignment model has a bimodal distribution, peaking at two angles that are close to the maximum values ($\pm 19.5^\circ$) expected between the elliptical and circular pattern for an inclination angle $i=45^\circ$ (see Fig. \ref{fig:angles}, lower panel). This is additional evidence that the polarization pattern observed in ALMA Band 3 is closer to elliptical than circular and that the radiative alignment model is disfavored based on the polarization orientation.  

Fig.~\ref{fig:maps} compares the 3 mm observations to the expected morphologies of polarized intensity for radiative alignment and aeroydynamic alignment. Both models poorly fit the observations.  The radiative alignment model produces more polarization at locations along the major axis than along the minor axis, and the opposite is true for the aerodynamic alignment model. The modeled patterns are in line with the expectations discussed in the last section based on the variation of the inclination of the grain alignment (symmetry) axis to the line of sight at different locations on the disk. Specifically, the higher polarization at locations along the major axis in the case of radiative alignment is because the effectively oblate grains there are viewed edge-on; those on the minor axis are viewed more face-on and thus appear rounder to the observer, yielding a lower polarization. In contrast, the effectively prolate grains in the case of aerodynamic alignment are viewed edge-on at locations along the minor axis (yieling maximum polarization) and more pole-on (and thus appear rounder) at locations along the major axis (yielding a lower polarization).  Beam averaging modifies the patterns somewhat, but not fundamentally. In particular, it does not average out the polarization near the center  because the polarization degree varies substantially in the azimuthal direction in both models, which contradicts the observation that shows a low-polarization ``hole" near the center (see Fig.~\ref{fig:observation}a). The strong discrepancy between the data and the models suggests that, by itself, neither radiative nor aerodynamic alignment explains well the observed data. In the next section, we will speculate on whether more complex models can explain the data better.

\subsection{More complex models: scattering by aerodynamically aligned grains?}
\label{sec:discussion}

%
In the previous section (\S~\ref{sec:data}), we have shown that the aerodynamic alignment model can explain the orientation of the polarization observed in HL Tau at Band 3 reasonably well (see Fig.~\ref{fig:hist}, top panel). However, it predicts a strong polarization parallel to the major axis at the center despite beam smearing and a lack of polarization at locations along the major axis, neither of which is observed. These two problems have the same origin: the decrease of polarization degree going from the minor axis to the major axis (see Fig.~\ref{fig:pang}). In order to make the model agree better with the data, one needs to find a way to increase the polarization degree for locations on the major axis relative to those on the minor axis without changing the polarization orientation. One natural way to meet this requirement, at least qualitatively, is through scattering. 

Previous studies have established that scattering in the Rayleigh limit produces a stronger polarization at locations on the major axis of an inclined disk than those on the minor axis (\citealt{Yang2016a}, see their Fig.~2; \citealt{Kataoka2016a}). Although the details of this azimuthal variation depend on the properties of the incident radiation, especially its anisotropy, it can be understood easily in the extreme case where most of the incident radiation is emitted by the brightest central region. In such a case, the incident radiation moving radially outward would be scattered by the grains located on the major axis by $90^\circ$ into the line of sight and thus be maximally polarized, but by an angle $90^\circ\pm i$ (where $i$ is the disk inclination angle) by the grains located on the minor axis in the disk plane, yielding a lower polarization. This tendency is broadly similar to that of the radiative alignment case, and the opposite of that of the aerodynamic alignment case. It is therefore reasonable to expect that when both direct emission and scattering by aerodynamically aligned grains are taken into account, the opposite tendencies for the polarization produced by direction emission and scattering should cancel each other at least to some extent, making the combined polarization less dependent on the azimuthal angle than that produced by direct emission alone. 

\begin{figure*}
\centering
\includegraphics[width=0.495\textwidth]{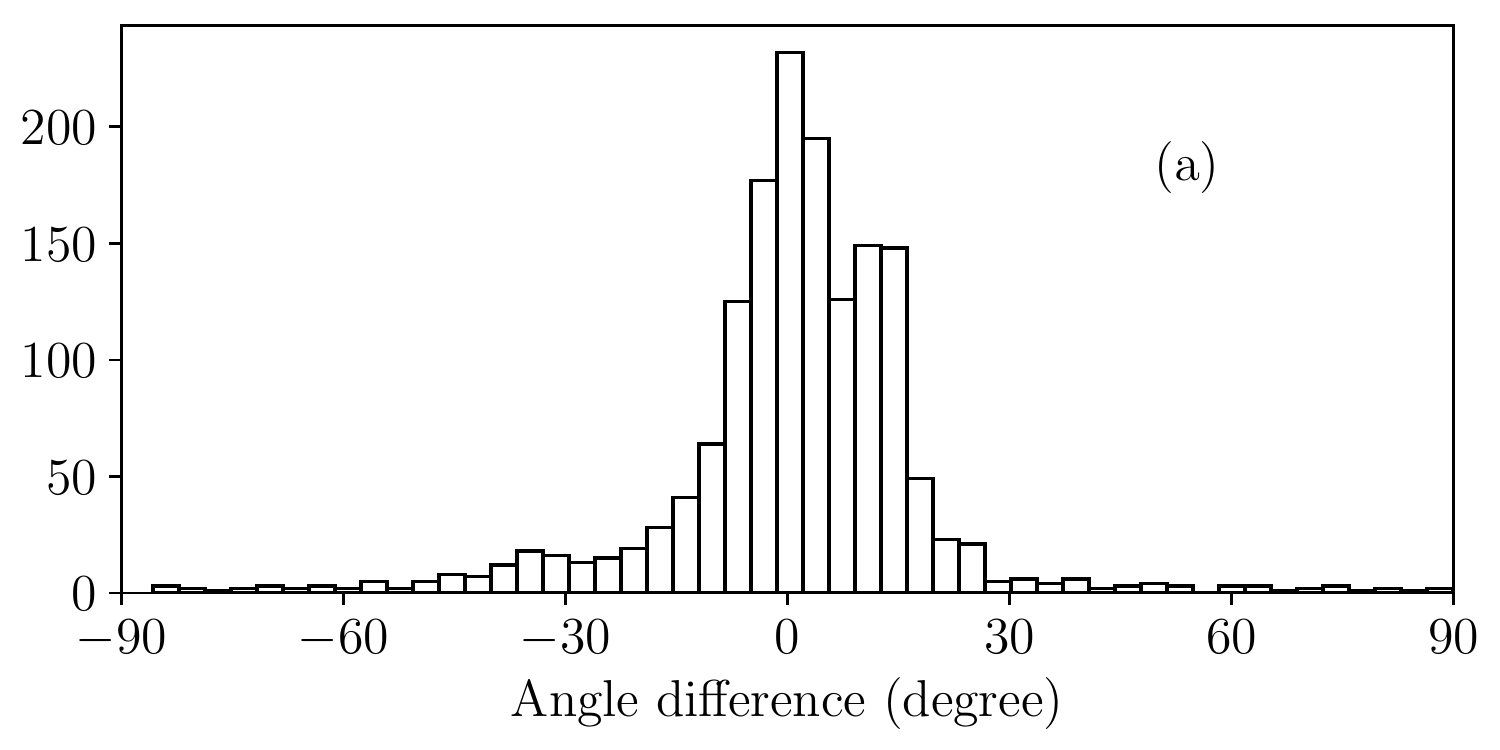}
\includegraphics[width=0.495\textwidth]{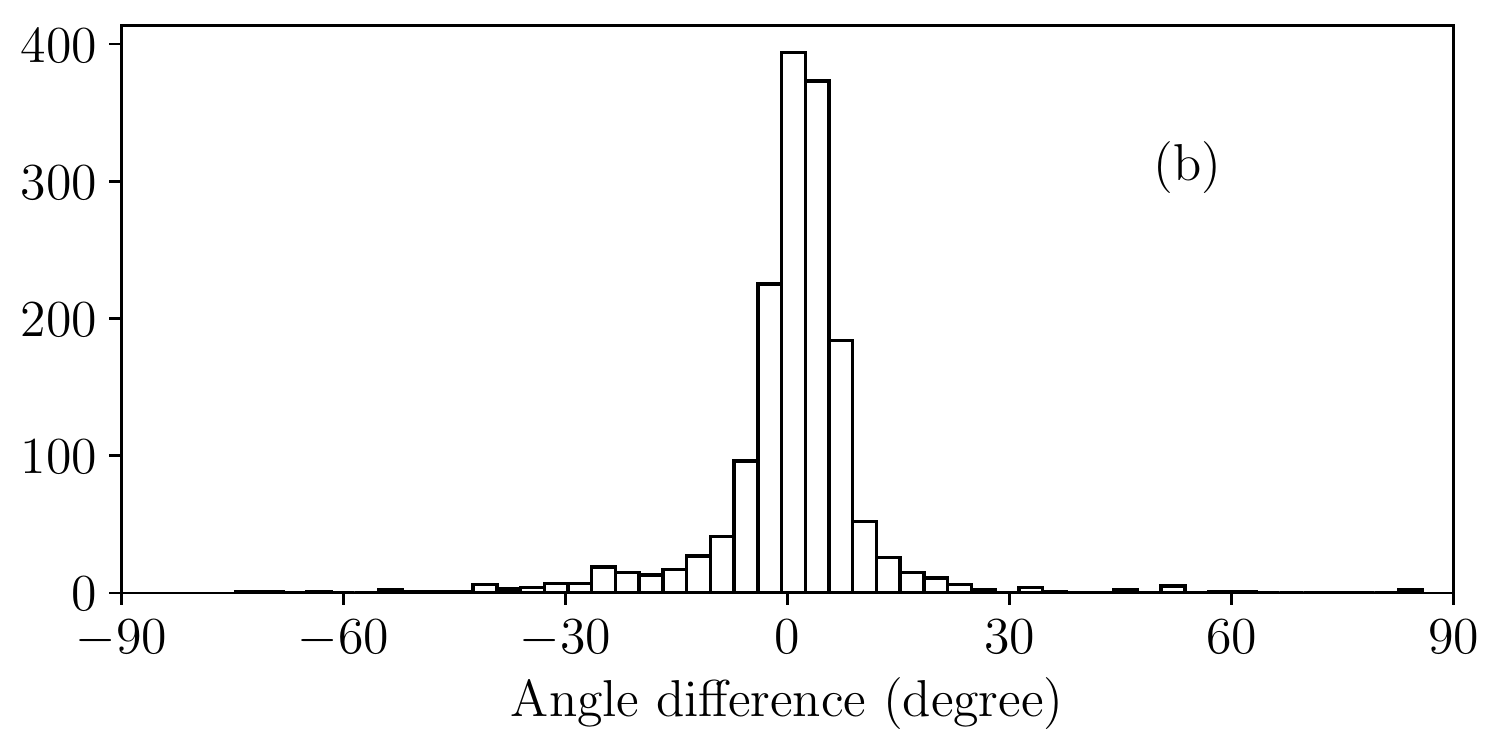}
\includegraphics[width=0.495\textwidth]{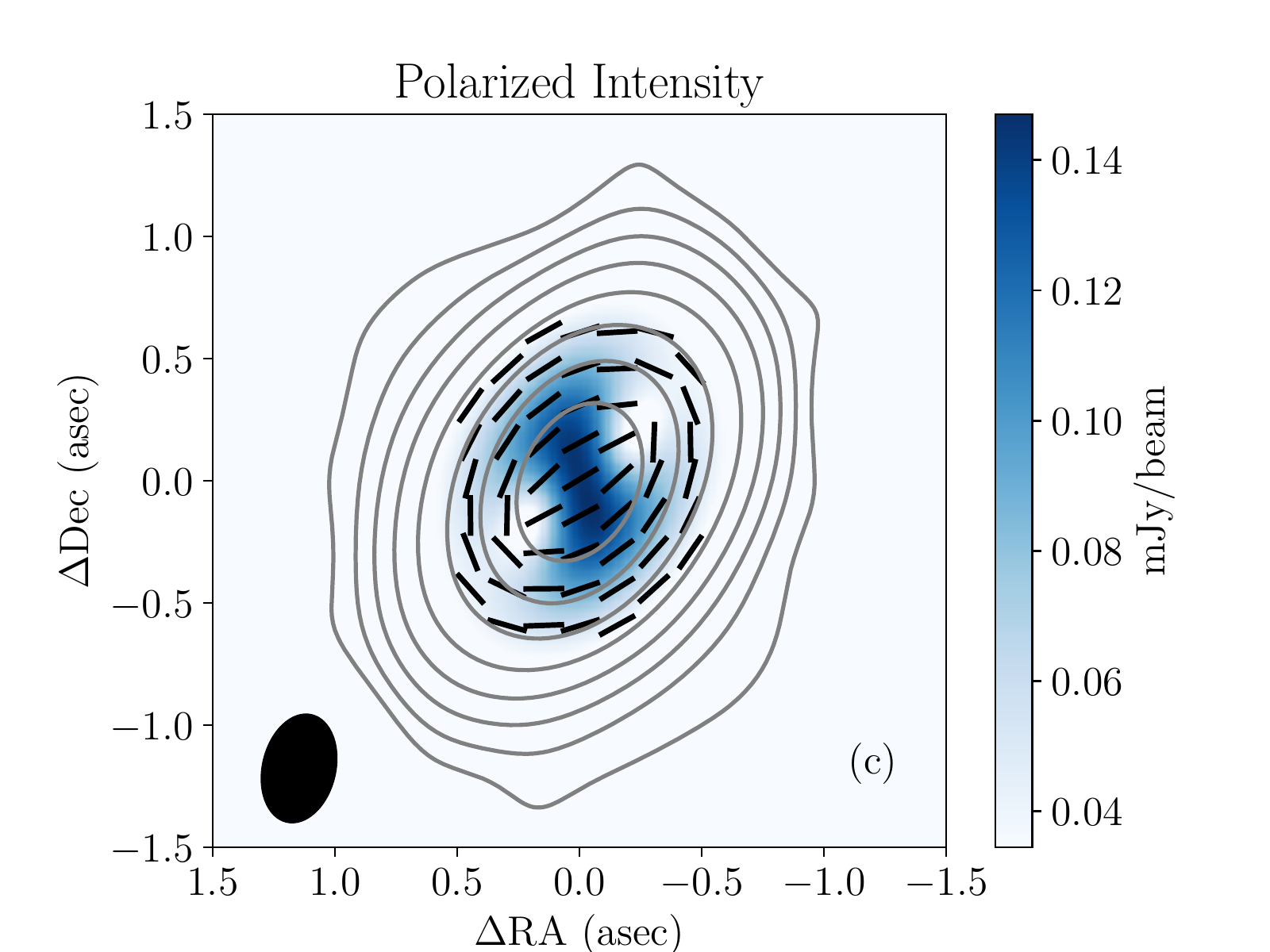}
\includegraphics[width=0.495\textwidth]{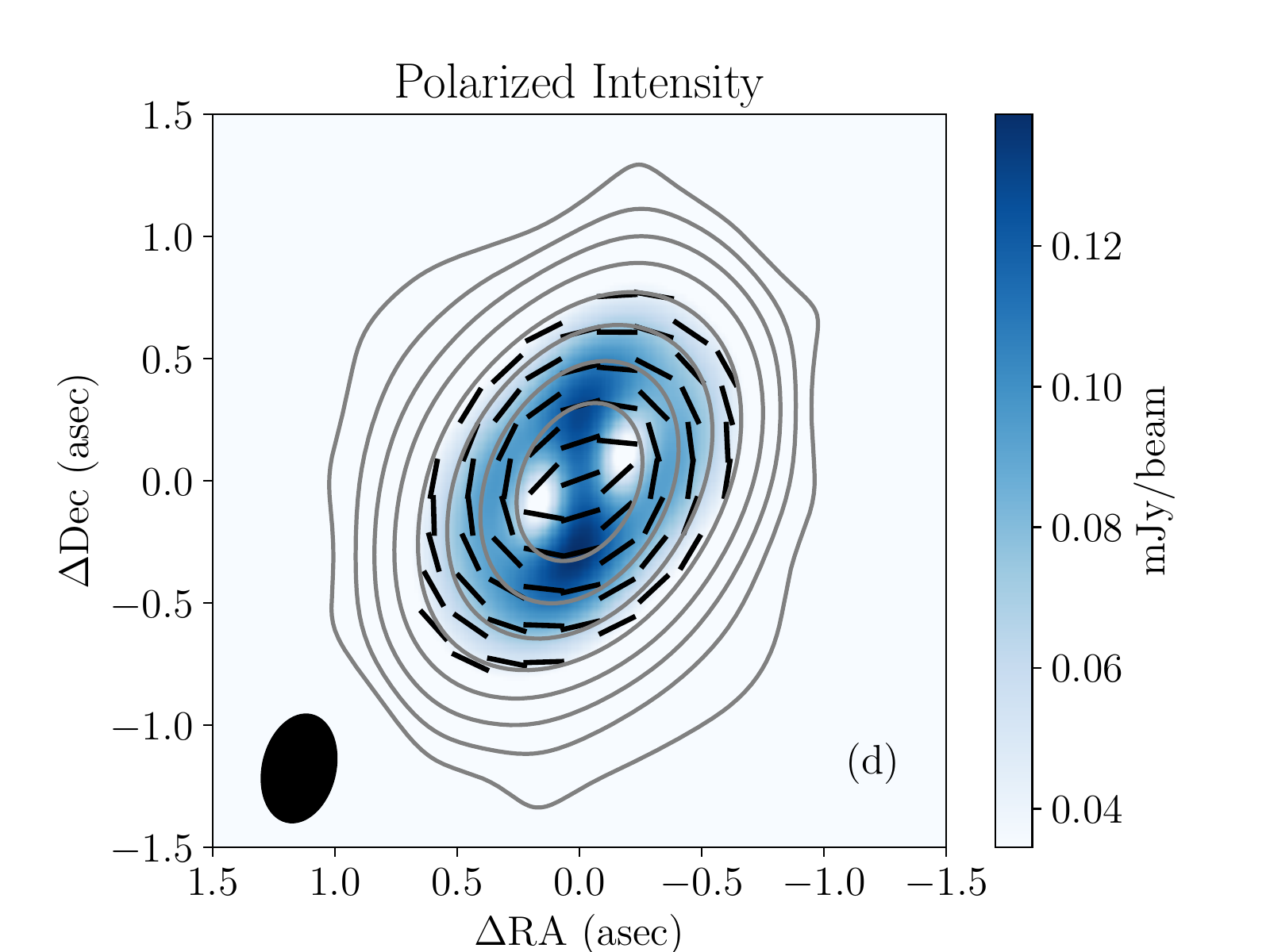}
\caption{Two intrinsically elliptical polarization models with different azimuthal variations. The left
column assumes uniform azimuthal polarization fraction, whereas the right column has 
polarization peaking at major axis, in the same way as the radiative alignment model. 
The upper panels show the histogram of the difference in polarization orientation with data
for the two models, and the lower panel show the simulated polarization observation.}
\label{fig:fake}
\end{figure*}

Whether the expected reduction in the degree of azimuthal variation of the polarization intensity can reproduce the observed data quantitatively remains to be determined. A self-consistent computation of the polarization from both direct emission and scattering by aerodynamically aligned grains is beyond the scope of this work. As an illustration of the basic principles, we carry out two numerical experiments. First, we reconsider the aerodynamic alignment model discussed in the last section, but with the azimuthal variation of the polarized intensity removed. We also normalize the polarization intensity so that the maximum value is close to the maximum observed value. The results are shown in Fig.~\ref{fig:fake}a,c. Compared to the unmodified aerodynamic alignment model of the last section, the beam-convolved polarization orientations agree with the observed values somewhat better (compare Fig.\ref{fig:fake}c to Fig.\ref{fig:maps}a), although there is still a substantial spread in their difference around $0^\circ$. Even though the polarized intensity is set to be azimuthally uniform intrinsically, it has pronounced azimuthal variation after beam-convolution. Specifically, there are two low polarization ``holes" located near the major axis (one on each side of the center). These are the regions where the polarization orientations in the intrinsically elliptical pattern change rapidly from one location to another. As a result, their polarized intensity is lowered more by beam-averaging compared to, e.g., the regions near the minor axis where the polarization orientations are more spatially uniform. This example demonstrates clearly the strong interplay between the spatial distribution of polarized intensity and the spatial variation of polarization orientation in the presence of significant beam-averaging; the effects of beam-convolution need to be evaluated carefully when comparing models and observations. 

To better match the observed 3 mm polarization pattern, the aerodynamic model would require an  intrinsic polarization that is higher along the major axis than along the minor axis. 
As an illustration, we adopt the azimuthal variation for the radiative alignment model for an inclination angle $i=45^\circ$ (shown in Fig.~\ref{fig:pang} as blue dashed line), where the polarization degree is about a factor of 2 higher on the major axis than on the minor axis. The results are shown Fig.~\ref{fig:fake}b,d. We have again normalized the maximum polarized intensity to the observed maximum value. Compared to the modified model with an intrinsically uniform azimuthal distribution of polarized intensity, there is a drastic improvement in the agreement between the modeled polarization orientations and the observed ones (compare Fig.~\ref{fig:fake}b to Fig.~\ref{fig:fake}a). The spatial distribution of polarized intensity also agrees better with observation, although a nearly vertical region of relatively high polarized intensity remains prominent near the center; it is barely visible in the data. The polarization near the center could in principle be reduced by a higher optical depth there, although the optical depth effects remain to be quantified. 

In any case, we have demonstrated that a combination of an intrinsic elliptical polarization pattern and an intrinsic azimuthal variation of polarization intensity that favors the major axis over the minor axis improves the model fit to the observed data in ALMA Band 3. Whether this combination can be achieved by direct emission and scattering by aerodynamically aligned grains or some other physical mechanisms remains to be determined. 

\subsection{Polarization spectrum: a potential problem for aligned grains?}
%


We have argued that it is difficult for the radiative alignment to explain the polarization observed in the HL Tau disk in ALMA Band 3 because it predicts a circular polarization pattern and substantial azimuthal variation of polarized intensity that are not observed. Another potential difficulty is that radiatively aligned grains are expected to at least contribute to, if not dominate, the polarization in other ALMA bands, especially Band 7 (see Fig.~\ref{fig:observation}b). 
If we assume a constant alignment efficiency and a single dust species, the polarization fraction changes little with wavelength, as long as the dielectric constants of the grains are well behaved, which is generally the case in the (sub)millimeter regime \citep{Draine1984,Kataoka2014}. If this is true, it would contradict the ALMA Band 7 polarization data on the HL Tau disk: the $\sim 1.8\%$ polarization detected in Band 3 \citep{Kataoka2017} is well above the $\sim 0.6\%$ polarization detected in Band 7 and has a completely different pattern \citep{Stephens2017}; the Band 7 polarization pattern is a textbook example of scattering-induced polarization in an inclined disk \citep{Yang2016a,Kataoka2016a}. The apparent lack of contamination from aligned grains in Band 7 posts a challenge to not only the radiative alignment mechanism but also aligned grain interpretation in general, including aerodynamically aligned grains. For aligned grain models to be compatible with the existing multi-wavelength data in HL Tau, their polarization fraction has to drop by at least a factor of 3 going from ALMA Band 3 ($\sim 3$~mm) to Band 7 ($\sim 0.87$~mm). 
A drop of polarization fraction with decreasing wavelength was predicted by \cite{Draine2009} in (sub)millimeter (see the right panel of their Fig.~8) for models with aligned silicate but not carbon grains. Whether these or other models can explain the required drop quantitatively remains to be determined.

\section{Summary and conclusions}
\label{sec:summary}


In this paper, we have discussed the polarization expected from different grain alignment
mechanisms, especially the radiative and aerodynamic alignment and compared the model predictions with the HL Tau ALMA Band 3 data. The main results are: 

\begin{enumerate}
    \item Unlike generally assumed previously, the polarization pattern from radiative alignment is circular (as in the optical polarization of reflection nebulae) rather than elliptical for an axisymmetric disk. The circular polarization expected of radiative alignment is not consistent with the pattern observed in the HL Tau disk in  ALMA Band 3 ($\sim 3$~mm), as shown in Fig.~\ref{fig:hist}b.
    \item An intrinsically elliptical pattern can be produced if the grains are aligned aerodynamically by the relative motions between the dust and gas in the azimuthal direction in the disk plane. The polarization orientations from the elliptical pattern are in better agreement with the Band 3 data, although a significant scatter remains because of beam-averaging in the simplest case where the polarization intensity does not have any intrinsic azimuthal variation (see Fig.~\ref{fig:fake}a).
    \item Strong intrinsic azimuthal variation is expected in an inclined disk for all grain alignment models, as shown in Fig. \ref{fig:pang}. In particular, the polarization is higher at locations on the minor axis for both the magnetic and aerodynamic alignment than those on the minor axis, and the opposite is true for the radiative alignment. The differences in both the polarization orientation and azimuthal variation of polarized intensity increase with the disk inclination angle to the line of sight, making edge-on disks ideally suited for distinguishing the different alignment mechanisms. 
    \item The strong azimuthal variation in polarized intensity expected in the radiative and aerodynamic alignment is not observed in the ALMA Band 3 polarization data of the HL Tau disk (see Fig.~\ref{fig:maps}), which is evidence against interpreting the data using either of these two mechanisms alone.
    Similar difficulties exist for grains aligned through mechanical alignment torques (MATs).
    \item We showed that beam-averaging introduces a strong coupling between the polarization orientation and azimuthal variation of the polarized intensity that needs to be accounted for when comparing models and data. In particular, a polarization pattern that is intrinsically elliptical without any intrinsic azimuthal variation in polarized intensity shows a pronounced azimuthal variation when beam-averaged (see Fig.~\ref{fig:fake}c). To reduce the azimuthal variation after beam-averaging (for a better match to observation), an intrinsic azimuthal variation with higher polarization along the major axis than along the minor axis is needed (see Fig.~\ref{fig:fake}b,d). Such an intrinsic variation is qualitatively expected of the polarization produced by dust scattering in an inclined disk. Whether a combination of both direct emission and scattering by aligned grains in general, and aerodynamically aligned grains in particular, can explain the ALMA Band 3 data remains to be determined. 
    \item We note that the polarization fraction
    detected in the HL Tau disk is significantly higher in Band 3 than Band 7 (by a factor of $\sim 3$ typically). Any grain alignment-based mechanism for explaining the Band 3 data will need to address the question of why the Band 7 data appears to be consistent with pure scattering, with little contamination from emission by aligned grains, which are expected to produce a polarization fraction that varies little with wavelength in the simplest dipole or electrostatic regime. More work is needed to resolve this potentially serious discrepancy. 
\end{enumerate}

We conclude that although the origin of the HL Tau disk polarization in ALMA Band 3 remains a mystery, the flood of ALMA data and relatively early stage of theoretical development should make the field of disk polarization an exciting area of research that is poised for rapid growth. 

\vspace{1em}
\textbf{ACKNOWLEDGEMENT}

We thank Dom Pesce for useful discussions. 
We thank an anonymous referee for thorough comments which helped improve 
our manuscript.
HY is supported in part by an NRAO ALMA SOS award. ZYL is supported in part by NSF grant AST-1815784 and AST-1716259 and NASA grant 80NSSC18K1095 and NNX14AB38G.

This paper makes use of the following ALMA
data: ADS/JAO.ALMA\#2011.0.00015.SV, 
ADS/JAO.ALMA\#2016.1.00115.S, and
ADS/JAO.ALMA\#2016.1.00162.S. ALMA is a partnership
of ESO (representing its member states), NSF
(USA) and NINS (Japan), together with NRC (Canada);
NSC and ASIAA (Taiwan); and KASI (Republic of
Korea), in cooperation with the Republic of Chile.
The Joint ALMA Observatory is operated by ESO,
AUI/NRAO and NAOJ. The National Radio Astronomy
Observatory is a facility of the National Science
Foundation operated under cooperative agreement by
Associated Universities, Inc.
This research made use of Astropy \citep{Astropy}
and matplotlib \citep{Matplotlib}.

 \newcommand{\noop}[1]{}

\label{lastpage}


\begin{thebibliography}{}
\makeatletter
\relax
\def\mn@urlcharsother{\let\do\@makeother \do\$\do\&\do\#\do\^\do\_\do\%\do\~}
\def\mn@doi{\begingroup\mn@urlcharsother \@ifnextchar [ {\mn@doi@}
  {\mn@doi@[]}}
\def\mn@doi@[#1]#2{\def\@tempa{#1}\ifx\@tempa\@empty \href
  {http://dx.doi.org/#2} {doi:#2}\else \href {http://dx.doi.org/#2} {#1}\fi
  \endgroup}
\def\mn@eprint#1#2{\mn@eprint@#1:#2::\@nil}
\def\mn@eprint@arXiv#1{\href {http://arxiv.org/abs/#1} {{\tt arXiv:#1}}}
\def\mn@eprint@dblp#1{\href {http://dblp.uni-trier.de/rec/bibtex/#1.xml}
  {dblp:#1}}
\def\mn@eprint@#1:#2:#3:#4\@nil{\def\@tempa {#1}\def\@tempb {#2}\def\@tempc
  {#3}\ifx \@tempc \@empty \let \@tempc \@tempb \let \@tempb \@tempa \fi \ifx
  \@tempb \@empty \def\@tempb {arXiv}\fi \@ifundefined
  {mn@eprint@\@tempb}{\@tempb:\@tempc}{\expandafter \expandafter \csname
  mn@eprint@\@tempb\endcsname \expandafter{\@tempc}}}

\bibitem[\protect\citeauthoryear{{ALMA Partnership} et~al.,}{{ALMA Partnership}
  et~al.}{2015}]{ALMA2015}
{ALMA Partnership} et~al., 2015, \apjl, 808, L3

\bibitem[\protect\citeauthoryear{{Alves} et~al.,}{{Alves}
  et~al.}{2018}]{Alves2018}
{Alves} F.~O.,  et~al., 2018, \mn@doi [\aap] {10.1051/0004-6361/201832935},
  \href {http://adsabs.harvard.edu/abs/2018A%26A...616A..56A} {616, A56}

\bibitem[\protect\citeauthoryear{Andersson, Lazarian  \&
  Vaillancourt}{Andersson et~al.}{2015}]{Andersson2015}
Andersson B.-G.,  Lazarian A.,   Vaillancourt J.~E.,  2015, \araa, 53, 501

\bibitem[\protect\citeauthoryear{{Balbus} \& {Hawley}}{{Balbus} \&
  {Hawley}}{1991}]{Balbus1991}
{Balbus} S.~A.,  {Hawley} J.~F.,  1991, \apj, 376, 214

\bibitem[\protect\citeauthoryear{{Bertrang} \& {Wolf}}{{Bertrang} \&
  {Wolf}}{2017}]{Bertrang2017}
{Bertrang} G.~H.-M.,  {Wolf} S.,  2017, \mn@doi [\mnras]
  {10.1093/mnras/stx1066}, \href
  {http://adsabs.harvard.edu/abs/2017MNRAS.469.2869B} {469, 2869}

\bibitem[\protect\citeauthoryear{Blandford \& Payne}{Blandford \&
  Payne}{1982}]{Blandford1982}
Blandford R.~D.,  Payne D.~G.,  1982, \mnras, 199, 883

\bibitem[\protect\citeauthoryear{Cho \& Lazarian}{Cho \&
  Lazarian}{2007}]{Cho2007}
Cho J.,  Lazarian A.,  2007, \apj, 669, 1085

\bibitem[\protect\citeauthoryear{{Cox}, {Harris}, {Looney}, {Li}, {Yang},
  {Tobin}  \& {Stephens}}{{Cox} et~al.}{2018}]{Cox2018}
{Cox} E.~G.,  {Harris} R.~J.,  {Looney} L.~W.,  {Li} Z.-Y.,  {Yang} H.,
  {Tobin} J.~J.,   {Stephens} I.,  2018, \mn@doi [\apj]
  {10.3847/1538-4357/aaacd2}, \href
  {http://adsabs.harvard.edu/abs/2018ApJ...855...92C} {855, 92}

\bibitem[\protect\citeauthoryear{{Davis} \& {Greenstein}}{{Davis} \&
  {Greenstein}}{1951}]{DG1951}
{Davis} Jr. L.,  {Greenstein} J.~L.,  1951, \mn@doi [\apj] {10.1086/145464},
  \href {http://adsabs.harvard.edu/abs/1951ApJ...114..206D} {114, 206}

\bibitem[\protect\citeauthoryear{{Dolginov} \& {Mitrofanov}}{{Dolginov} \&
  {Mitrofanov}}{1976}]{Dolginov1976}
{Dolginov} A.~Z.,  {Mitrofanov} I.~G.,  1976, \mn@doi [\apss]
  {10.1007/BF00640010}, \href
  {http://adsabs.harvard.edu/abs/1976Ap%26SS..43..291D} {43, 291}

\bibitem[\protect\citeauthoryear{{Draine} \& {Fraisse}}{{Draine} \&
  {Fraisse}}{2009}]{Draine2009}
{Draine} B.~T.,  {Fraisse} A.~A.,  2009, \mn@doi [\apj]
  {10.1088/0004-637X/696/1/1}, \href
  {http://adsabs.harvard.edu/abs/2009ApJ...696....1D} {696, 1}

\bibitem[\protect\citeauthoryear{{Draine} \& {Lee}}{{Draine} \&
  {Lee}}{1984}]{Draine1984}
{Draine} B.~T.,  {Lee} H.~M.,  1984, \mn@doi [\apj] {10.1086/162480}, \href
  {http://adsabs.harvard.edu/abs/1984ApJ...285...89D} {285, 89}

\bibitem[\protect\citeauthoryear{{Fissel} et~al.,}{{Fissel}
  et~al.}{2016}]{Fissel2016}
{Fissel} L.~M.,  et~al., 2016, \mn@doi [\apj] {10.3847/0004-637X/824/2/134},
  \href {http://adsabs.harvard.edu/abs/2016ApJ...824..134F} {824, 134}

\bibitem[\protect\citeauthoryear{{Girart}, {Rao}  \& {Marrone}}{{Girart}
  et~al.}{2006}]{Girart2006}
{Girart} J.~M.,  {Rao} R.,   {Marrone} D.~P.,  2006, \mn@doi [Science]
  {10.1126/science.1129093}, \href
  {http://adsabs.harvard.edu/abs/2006Sci...313..812G} {313, 812}

\bibitem[\protect\citeauthoryear{{Girart} et~al.,}{{Girart}
  et~al.}{2018}]{Girart2018}
{Girart} J.~M.,  et~al., 2018, \mn@doi [\apjl] {10.3847/2041-8213/aab76b},
  \href {http://adsabs.harvard.edu/abs/2018ApJ...856L..27G} {856, L27}

\bibitem[\protect\citeauthoryear{{Gold}}{{Gold}}{1952}]{Gold1952}
{Gold} T.,  1952, \mn@doi [\mnras] {10.1093/mnras/112.2.215}, \href
  {http://adsabs.harvard.edu/abs/1952MNRAS.112..215G} {112, 215}

\bibitem[\protect\citeauthoryear{{Hoang}}{{Hoang}}{2017}]{Hoang2017}
{Hoang} T.,  2017, preprint, \href
  {http://adsabs.harvard.edu/abs/2017arXiv170401721H} {} (\mn@eprint {arXiv}
  {1704.01721})

\bibitem[\protect\citeauthoryear{{Hoang} \& {Lazarian}}{{Hoang} \&
  {Lazarian}}{2014}]{Hoang2014}
{Hoang} T.,  {Lazarian} A.,  2014, \mn@doi [\mnras] {10.1093/mnras/stt2240},
  \href {http://adsabs.harvard.edu/abs/2014MNRAS.438..680H} {438, 680}

\bibitem[\protect\citeauthoryear{{Hoang} \& {Lazarian}}{{Hoang} \&
  {Lazarian}}{2016}]{HoangLazarian2016}
{Hoang} T.,  {Lazarian} A.,  2016, \mn@doi [\apj]
  {10.3847/0004-637X/831/2/159}, \href
  {http://adsabs.harvard.edu/abs/2016ApJ...831..159H} {831, 159}

\bibitem[\protect\citeauthoryear{{Hoang}, {Cho}  \& {Lazarian}}{{Hoang}
  et~al.}{2018}]{Hoang2018}
{Hoang} T.,  {Cho} J.,   {Lazarian} A.,  2018, \mn@doi [\apj]
  {10.3847/1538-4357/aa9edc}, \href
  {http://adsabs.harvard.edu/abs/2018ApJ...852..129H} {852, 129}

\bibitem[\protect\citeauthoryear{{Hull} et~al.,}{{Hull}
  et~al.}{2014}]{Hull2014}
{Hull} C.~L.~H.,  et~al., 2014, \mn@doi [\apjs] {10.1088/0067-0049/213/1/13},
  \href {http://adsabs.harvard.edu/abs/2014ApJS..213...13H} {213, 13}

\bibitem[\protect\citeauthoryear{{Hull} et~al.,}{{Hull}
  et~al.}{2018}]{Hull2018}
{Hull} C.~L.~H.,  et~al., 2018, \mn@doi [\apj] {10.3847/1538-4357/aabfeb},
  \href {http://adsabs.harvard.edu/abs/2018ApJ...860...82H} {860, 82}

\bibitem[\protect\citeauthoryear{Hunter}{Hunter}{2007}]{Matplotlib}
Hunter J.~D.,  2007, \mn@doi [Computing In Science \& Engineering]
  {10.1109/MCSE.2007.55}, 9, 90

\bibitem[\protect\citeauthoryear{{Jones} \& {Spitzer}}{{Jones} \&
  {Spitzer}}{1967}]{JonesSpitzer1967}
{Jones} R.~V.,  {Spitzer} Jr. L.,  1967, \mn@doi [\apj] {10.1086/149086}, \href
  {http://adsabs.harvard.edu/abs/1967ApJ...147..943J} {147, 943}

\bibitem[\protect\citeauthoryear{{Kataoka}, {Okuzumi}, {Tanaka}  \&
  {Nomura}}{{Kataoka} et~al.}{2014}]{Kataoka2014}
{Kataoka} A.,  {Okuzumi} S.,  {Tanaka} H.,   {Nomura} H.,  2014, \mn@doi [\aap]
  {10.1051/0004-6361/201323199}, \href
  {http://adsabs.harvard.edu/abs/2014A%26A...568A..42K} {568, A42}

\bibitem[\protect\citeauthoryear{Kataoka et~al.,}{Kataoka
  et~al.}{2015}]{Kataoka2015}
Kataoka A.,  et~al., 2015, \apj, 809, 78

\bibitem[\protect\citeauthoryear{Kataoka, Muto, Momose, Tsukagoshi  \&
  Dullemond}{Kataoka et~al.}{2016}]{Kataoka2016a}
Kataoka A.,  Muto T.,  Momose M.,  Tsukagoshi T.,   Dullemond C.~P.,  2016,
  \apj, 820, 54

\bibitem[\protect\citeauthoryear{{Kataoka}, {Tsukagoshi}, {Pohl}, {Muto},
  {Nagai}, {Stephens}, {Tomisaka}  \& {Momose}}{{Kataoka}
  et~al.}{2017}]{Kataoka2017}
{Kataoka} A.,  {Tsukagoshi} T.,  {Pohl} A.,  {Muto} T.,  {Nagai} H.,
  {Stephens} I.~W.,  {Tomisaka} K.,   {Momose} M.,  2017, \mn@doi [\apjl]
  {10.3847/2041-8213/aa7e33}, \href
  {http://adsabs.harvard.edu/abs/2017ApJ...844L...5K} {844, L5}

\bibitem[\protect\citeauthoryear{{Lazarian}}{{Lazarian}}{1995}]{Lazarian1995}
{Lazarian} A.,  1995, \mn@doi [\apj] {10.1086/176252}, \href
  {http://adsabs.harvard.edu/abs/1995ApJ...451..660L} {451, 660}

\bibitem[\protect\citeauthoryear{Lazarian}{Lazarian}{2007}]{Lazarian2007}
Lazarian A.,  2007, \mn@doi [Journal of Quantitative Spectroscopy and Radiative
  Transfer] {https://doi.org/10.1016/j.jqsrt.2007.01.038}, 106, 225

\bibitem[\protect\citeauthoryear{{Lazarian} \& {Hoang}}{{Lazarian} \&
  {Hoang}}{2007a}]{LH2007}
{Lazarian} A.,  {Hoang} T.,  2007a, \mn@doi [\mnras]
  {10.1111/j.1365-2966.2007.11817.x}, \href
  {http://adsabs.harvard.edu/abs/2007MNRAS.378..910L} {378, 910}

\bibitem[\protect\citeauthoryear{{Lazarian} \& {Hoang}}{{Lazarian} \&
  {Hoang}}{2007b}]{LazarianHoang2017M}
{Lazarian} A.,  {Hoang} T.,  2007b, \mn@doi [\apjl] {10.1086/523849}, \href
  {http://adsabs.harvard.edu/abs/2007ApJ...669L..77L} {669, L77}

\bibitem[\protect\citeauthoryear{{Lee} \& {Draine}}{{Lee} \&
  {Draine}}{1985}]{LeeDraine1985}
{Lee} H.~M.,  {Draine} B.~T.,  1985, \mn@doi [\apj] {10.1086/162974}, \href
  {http://adsabs.harvard.edu/abs/1985ApJ...290..211L} {290, 211}

\bibitem[\protect\citeauthoryear{{Lee}, {Li}, {Ching}, {Lai}  \& {Yang}}{{Lee}
  et~al.}{2018}]{Lee2018}
{Lee} C.-F.,  {Li} Z.-Y.,  {Ching} T.-C.,  {Lai} S.-P.,   {Yang} H.,  2018,
  \mn@doi [\apj] {10.3847/1538-4357/aaa769}, \href
  {http://adsabs.harvard.edu/abs/2018ApJ...854...56L} {854, 56}

\bibitem[\protect\citeauthoryear{{Matsakos}, {Tzeferacos}  \&
  {K{\"o}nigl}}{{Matsakos} et~al.}{2016}]{Matsakos2016}
{Matsakos} T.,  {Tzeferacos} P.,   {K{\"o}nigl} A.,  2016, \mn@doi [\mnras]
  {10.1093/mnras/stw2175}, \href
  {http://adsabs.harvard.edu/abs/2016MNRAS.463.2716M} {463, 2716}

\bibitem[\protect\citeauthoryear{{Planck Collaboration} et~al.,}{{Planck
  Collaboration} et~al.}{2015}]{PlanckXIX2014}
{Planck Collaboration} et~al., 2015, \mn@doi [\aap]
  {10.1051/0004-6361/201424082}, \href
  {http://adsabs.harvard.edu/abs/2015A%26A...576A.104P} {576, A104}

\bibitem[\protect\citeauthoryear{{Purcell}}{{Purcell}}{1979}]{Purcell1979}
{Purcell} E.~M.,  1979, \mn@doi [\apj] {10.1086/157204}, \href
  {http://adsabs.harvard.edu/abs/1979ApJ...231..404P} {231, 404}

\bibitem[\protect\citeauthoryear{{Stephens} et~al.,}{{Stephens}
  et~al.}{2013}]{Stephens2013}
{Stephens} I.~W.,  et~al., 2013, \mn@doi [\apjl] {10.1088/2041-8205/769/1/L15},
  \href {http://adsabs.harvard.edu/abs/2013ApJ...769L..15S} {769, L15}

\bibitem[\protect\citeauthoryear{Stephens et~al.,}{Stephens
  et~al.}{2014}]{Stephens2014}
Stephens I.~W.,  et~al., 2014, \nat, 514, 597

\bibitem[\protect\citeauthoryear{{Stephens} et~al.,}{{Stephens}
  et~al.}{2017}]{Stephens2017}
{Stephens} I.~W.,  et~al., 2017, \mn@doi [\apj] {10.3847/1538-4357/aa998b},
  \href {http://adsabs.harvard.edu/abs/2017ApJ...851...55S} {851, 55}

\bibitem[\protect\citeauthoryear{{Tazaki}, {Lazarian}  \& {Nomura}}{{Tazaki}
  et~al.}{2017}]{Tazaki2017}
{Tazaki} R.,  {Lazarian} A.,   {Nomura} H.,  2017, \mn@doi [\apj]
  {10.3847/1538-4357/839/1/56}, \href
  {http://adsabs.harvard.edu/abs/2017ApJ...839...56T} {839, 56}

\bibitem[\protect\citeauthoryear{{The Astropy Collaboration} et~al.,}{{The
  Astropy Collaboration} et~al.}{2018}]{Astropy}
{The Astropy Collaboration} et~al., 2018, preprint, \href
  {http://adsabs.harvard.edu/abs/2018arXiv180102634T} {} (\mn@eprint {arXiv}
  {1801.02634})

\bibitem[\protect\citeauthoryear{Turner, Fromang, Gammie, Klahr, Lesur, Wardle
  \& Bai}{Turner et~al.}{2014}]{Turner2014}
Turner N.~J.,  Fromang S.,  Gammie C.,  Klahr H.,  Lesur G.,  Wardle M.,   Bai
  X.~N.,  2014, \mn@doi [Protostars and Planets VI, Henrik Beuther, Ralf S.
  Klessen, Cornelis P. Dullemond, and Thomas Henning (eds.), University of
  Arizona Press, Tucson] {10.2458/azu_uapress_9780816531240-ch018}, pp 411--432

\bibitem[\protect\citeauthoryear{Yang, Li, Looney  \& Stephens}{Yang
  et~al.}{2016a}]{Yang2016a}
Yang H.,  Li Z.-Y.,  Looney L.,   Stephens I.,  2016a, \mnras, 456, 2794

\bibitem[\protect\citeauthoryear{Yang, Li, Looney, Cox, Tobin, Stephens,
  Segura-Cox  \& Harris}{Yang et~al.}{2016b}]{Yang2016b}
Yang H.,  Li Z.-Y.,  Looney L.~W.,  Cox E.~G.,  Tobin J.,  Stephens I.~W.,
  Segura-Cox D.~M.,   Harris R.~J.,  2016b, \mnras, 460, 4109

\bibitem[\protect\citeauthoryear{{Yang}, {Li}, {Looney}, {Girart}  \&
  {Stephens}}{{Yang} et~al.}{2017}]{Yang2017}
{Yang} H.,  {Li} Z.-Y.,  {Looney} L.~W.,  {Girart} J.~M.,   {Stephens} I.~W.,
  2017, \mn@doi [\mnras] {10.1093/mnras/stx1951}, \href
  {http://adsabs.harvard.edu/abs/2017MNRAS.472..373Y} {472, 373}

\makeatother
\end{thebibliography}
\end{document}